\documentclass[final]{raa06}           
\usepackage{graphicx,times}             
\usepackage{natbib}
\usepackage{amssymb,amsmath}
\bibpunct{(}{)}{;}{a}{}{,}
\usepackage[colorlinks=true, citecolor=blue]{hyperref}%

\usepackage{graphicx,kantlipsum,setspace}
\usepackage{caption}
\usepackage{graphics,epsf}
\usepackage{amsmath}                
\usepackage{amsfonts}               
\usepackage{amssymb}                
\usepackage{epsfig}                 
\usepackage{appendix}
\usepackage{graphicx}
\usepackage{float}
\usepackage{color}
\usepackage{multirow}
\usepackage{colortbl}
\usepackage[para,online,flushleft]{threeparttable}
\usepackage{xcolor}

\hypersetup{citecolor=blue, 
            linkcolor=red, 
            menucolor=blue, 
            urlcolor=blue}  

\newcommand{\cm}{{~\rm cm}}

\newcommand{\km}{{~\rm km}}
\newcommand{\s}{{~\rm s}}

\newcommand{\g}{{~\rm g}}

\newcommand{\erg}{{~\rm erg}}
\newcommand{\yr}{{~\rm yr}}

\newcommand{\keV}{{~\rm keV}}

\begin{document}

   \title{Review: The role of jets in exploding supernovae and in shaping their remnants
}

   \volnopage{Vol.0 (20xx) No.0, 000--000}      
   \setcounter{page}{1}          

   \author{Noam Soker
      \inst{1}
   }

   \institute{Department of Physics, Technion, Haifa, 3200003, Israel;   {\it soker@physics.technion.ac.il}\\
\vs\no
   {\small Received~~20xx month day; accepted~~20xx~~month day}}

\abstract{
I review studies of core collapse supernovae (CCSNe) and similar transient events that attribute major roles to jets in powering most CCSNe and in shaping their ejecta. I start with reviewing the jittering jets explosion mechanism that I take to power most CCSN explosions. Neutrino heating does play a role in boosting the jets. I compare the morphologies of some CCSN remnants to planetary nebulae to conclude that jets and instabilities are behind the shaping of their ejecta. I then discuss CCSNe that are descendants of rapidly rotating collapsing cores that result in fixed-axis jets (with small jittering) that shape bipolar ejecta. A large fraction of the bipolar CCSNe are superluminous supernovae (SLSNe). I conclude that modelling of SLSNe lightcurves and bumps in the lightcurves must include jets, even when considering energetic magnetars and/or ejecta interaction with the circumstellar matter (CSM). I connect the properties of bipolar CCSNe to common envelope jets supernovae (CEJSNe) where an old neutron star or a black hole spirals-in inside the envelope and then inside the core of a red supergiant. I discuss how jets can shape the pre-explosion CSM, as in supernova 1987A, and can power pre-explosion outbursts (precursors) in binary systems progenitors of CCSNe and CEJSNe. Binary interaction facilitate also the launching of post-explosion jets.
}

 \authorrunning{N. Soker}            
\titlerunning{The role of jets in supernovae}  
   
      \maketitle


\section{Introduction} 
\label{sec:intro}
In this review I summarize many, but not all, of the studies in the last decade that attribute major roles to jets in core collapse supernovae (CCSNe) and related energetic stellar transients, in particular common envelope jets supernovae (CEJSNe). I concentrate on studies that adopt the view that jets explode most, or even all, CCSNe. 
 
CEJSNe are defined as jet-powered energetic transient events where an old neutron star (NS) or an old black hole (BH) accrete mass from the envelope and then from the core of a red supergiant (RSG) star and launch energetic jets (section \ref{sec:CEJSNe}).  
I consider the role of jets in powering and in shaping events that are true CCSNe (sections \ref{sec:CCSNeJittering} and \ref{sec:CCSNeFixed}), CEJSNe, which during the first weeks of the event might be classified as CCSNe (section \ref{sec:CEJSNe}), pre-explosion shaping of the CSM in CCSNe and type Ia supernovae (SNe Ia; section \ref{sec:PreExplosionCSM}), pre-explosion outbursts (section \ref{sec:PreExplosionOutbursts}), and post-explosion powering (section \ref{sec:PostExplosion}).

The different sections reveal the very large parameter space that jets introduce. The parameter space includes regular CCSNe, as well as rare combinations of parameters that can explain peculiar transients. 

In this review I do not discuss high-energy processes inside the jets (gamma rays, cosmic ray acceleration, neutrino production, r-process nucleosynthesis) as each one of these processes requires a separate review. These processes might definitely take place in many of the systems with energetic jets that I review. 

Because the different sections review different scenarios and different processes, although in all of them jets play the major roles, this review has no summary of the results, but rather a summary of some open questions (section \ref{sec:Summary}). Instead, each section has its own short summary that together with Table \ref{TAB:Table1} serve as the summary of the review. 
%
\begin{table*}[]
\begin{tabular}{|p{1.8cm}|p{2.1cm}|p{1.3cm}|p{1.7cm}|p{2cm}|p{2cm}|p{2cm}|p{0.5cm}|}
\hline
\textbf{Phase} & \textbf{Compact object} & \textbf{System} & \textbf{Source of mass} & \textbf{Source of A.M.} & \textbf{Fraction} & \textbf{Event} & \textbf{\S} \\ \hline
{\color[HTML]{7030A0} Pre-Explosion} & {\color[HTML]{7030A0} MS} & {\color[HTML]{7030A0} Binary} & {\color[HTML]{7030A0} RSG/WR} & {\color[HTML]{7030A0} Orbital} & {\color[HTML]{7030A0} Small} & {\color[HTML]{7030A0} ILOT (LRN)} & {\color[HTML]{7030A0} } \ref{sec:PreExplosionCSM},\ref{sec:PreExplosionOutbursts} \\ \hline
{\color[HTML]{7030A0} Pre-Explosion} & {\color[HTML]{7030A0} NS/BH} & {\color[HTML]{7030A0} Binary} & {\color[HTML]{7030A0} RSG/WR} & {\color[HTML]{7030A0} Orbital} & {\color[HTML]{7030A0} Small} & {\color[HTML]{7030A0} CEJSN-impostor} & {\color[HTML]{7030A0} } \ref{sec:PreExplosionCSM}, \ref{sec:PreExplosionOutbursts} \\ \hline
{\color[HTML]{7030A0} Pre-Explosion} & {\color[HTML]{7030A0} 2(NS/BH/MS)} & {\color[HTML]{7030A0} Triple} & {\color[HTML]{7030A0} RSG/WR} & {\color[HTML]{7030A0} Orbital: triple   +inner binary} & {\color[HTML]{7030A0} Very Small} & {\color[HTML]{7030A0} CEJSN-impostor   (or ILOT for MSs)} & {\color[HTML]{7030A0} } \ref{sec:PreExplosionCSM}, \ref{sec:PreExplosionOutbursts} \\ \hline
{\color[HTML]{CC0033} Explosion} & {\color[HTML]{CC0033} New  NS} & {\color[HTML]{CC0033} S/B/T} & {\color[HTML]{CC0033} core} & {\color[HTML]{CC0033} Core   Convection  (jittering jets)} & {\color[HTML]{CC0033} Most CCSNe} & {\color[HTML]{CC0033} CCSN} & {\color[HTML]{CC0033} } \ref{sec:CCSNeJittering} \\ \hline
{\color[HTML]{CC0033} Explosion} & {\color[HTML]{CC0033} New NS} & {\color[HTML]{CC0033} B/T} & {\color[HTML]{CC0033} core} & {\color[HTML]{CC0033} Core rotation} & {\color[HTML]{CC0033} Small} & {\color[HTML]{CC0033} CCSN: SNe/SLSNe} & {\color[HTML]{CC0033} } \ref{sec:CCSNeFixed}, \ref{subsec:LongLastingjets}\\ \hline
{\color[HTML]{CC0033} Explosion} & {\color[HTML]{CC0033} New BH} & {\color[HTML]{CC0033} S/B/T} & {\color[HTML]{CC0033} Core + envelope} & {\color[HTML]{CC0033} Envelope   Convection (jittering jets)} & {\color[HTML]{CC0033} Small} & {\color[HTML]{CC0033} CCSN: SLSNe +peculiar CCSNe} & {\color[HTML]{CC0033} } \ref{sec:CCSNeJittering} \\ \hline
{\color[HTML]{CC0033} Explosion} & {\color[HTML]{CC0033} New BH} & {\color[HTML]{CC0033} B/T} & {\color[HTML]{CC0033} Core + Envelope} & {\color[HTML]{CC0033} Core + Envelope   rotation} & {\color[HTML]{CC0033} Medium} & {\color[HTML]{CC0033} CCSN: SLSNe +   peculiar CCSNe} & {\color[HTML]{CC0033} } \ref{sec:CCSNeFixed}, \ref{subsec:LongLastingjets} \\ \hline
{\color[HTML]{385723} Explosion} & {\color[HTML]{385723} Old NS/BH} & {\color[HTML]{385723} B/T} & {\color[HTML]{385723} Core in a CEE} & {\color[HTML]{385723} Orbital} & {\color[HTML]{385723} Small} & {\color[HTML]{385723} CEJSN (including   FBOTs)} & {\color[HTML]{385723} } \ref{sec:CEJSNe} \\ \hline
{\color[HTML]{0070C0} Post-Explosion} & {\color[HTML]{0070C0} New NS/BH} & {\color[HTML]{0070C0} S/B/T} & {\color[HTML]{0070C0} Fallback Ejecta} & {\color[HTML]{0070C0} Pre- Explosion rotation} & {\color[HTML]{0070C0} Small} & {\color[HTML]{0070C0} Extended lightcurve and/or bumps} & {\color[HTML]{0070C0} } \ref{sec:PostExplosion} \\ \hline
{\color[HTML]{0070C0} Post-Explosion} & {\color[HTML]{0070C0} New NS/BH} & {\color[HTML]{0070C0} Binary} & {\color[HTML]{0070C0} MS Companion} & {\color[HTML]{0070C0} Orbital} & {\color[HTML]{0070C0} Very small} & {\color[HTML]{0070C0} Extended   lightcurve and/or bumps} & {\color[HTML]{0070C0} } \ref{sec:PostExplosion} \\ \hline
{\color[HTML]{0070C0} Post-Explosion} & {\color[HTML]{0070C0} Old NS/BH} & {\color[HTML]{0070C0} Binary} & {\color[HTML]{0070C0} Ejecta} & {\color[HTML]{0070C0} Orbital} & {\color[HTML]{0070C0} Very small} & {\color[HTML]{0070C0} Small extra   energy} & {\color[HTML]{0070C0} } \ref{sec:CEJSNe}, \ref{sec:PostExplosion} \\ \hline
\end{tabular}
\caption{The main type of jet-driven events (seventh column) and the phase of the jet activity (first column) that I review in the different sections (last column). In the second column I list the compact object that accretes mass. In the third column I list the required type of interacting system. Note that single-star (S) processes can take place also in binary (B) systems and in triple (T) systems, and that binary processes might take place in triple-star systems as well. The fourth and fifth columns list the source of the accreted mass and its angular momentum with respect to the accreting object, respectively. The sixth column lists the fraction of such systems relative to the total number of CCSNe. Abbreviation. A.M.: angular momentum; BH: black hole; CCSNe: core collapse supernovae; CEE: common envelope evolution; CEJSN: common envelope jets supernova; FBOT: Fast blue optical transient; ILOT: intermediate luminosity optical transient (other names include red nova and luminous red nova, LRN); MS: Main sequence; NS: neutron star; RSG: red supergiant; S/B/T: single/binary/triple; SLSNe: superluminous supernovae; WR: Wolf–Rayet.  
}
\label{TAB:Table1}
\end{table*}
\normalsize

In Table \ref{TAB:Table1} I list the main jet-activity phases (first column) and transient types (next to last column). The table does not cover all possibilities. 
In the second column I list the compact object that accretes mass and launches the jets, a main sequence star (MS), a NS or a BH. In the third column I indicate whether the system might be a single star (S), a binary system (T), or a triple system (T). In the forth and fifth columns I indicate the main mass and angular momentum sources of the accreted mass, respectively. In the sixth column I list my crude estimate of the fraction of these events and phases.  I elaborate on each of these in the respective section that I list in the last column. The rich spectrum of processes and properties of jets explains not only regular CCSNe but also peculiar CCSNe and similar transients. 

\textbf{$\bigstar$ Summary of section \ref{sec:intro}.} There is a rich spectrum of processes by which jets can influence the properties of CCSNe and CEJSNe, including their pre-explosion, explosion, and post explosion shaping and light curves, e.g., bumps in the light curve (Table \ref{TAB:Table1}).  

\section{The jittering jets explosion mechanism} 
\label{sec:CCSNeJittering}

\subsection{The definition of the mechanism} 
\label{subsec:TheMechanism}

\begin{figure}[!t]
\includegraphics[trim=3.0cm 8.3cm 0.0cm 3.6cm ,clip, scale=0.51]{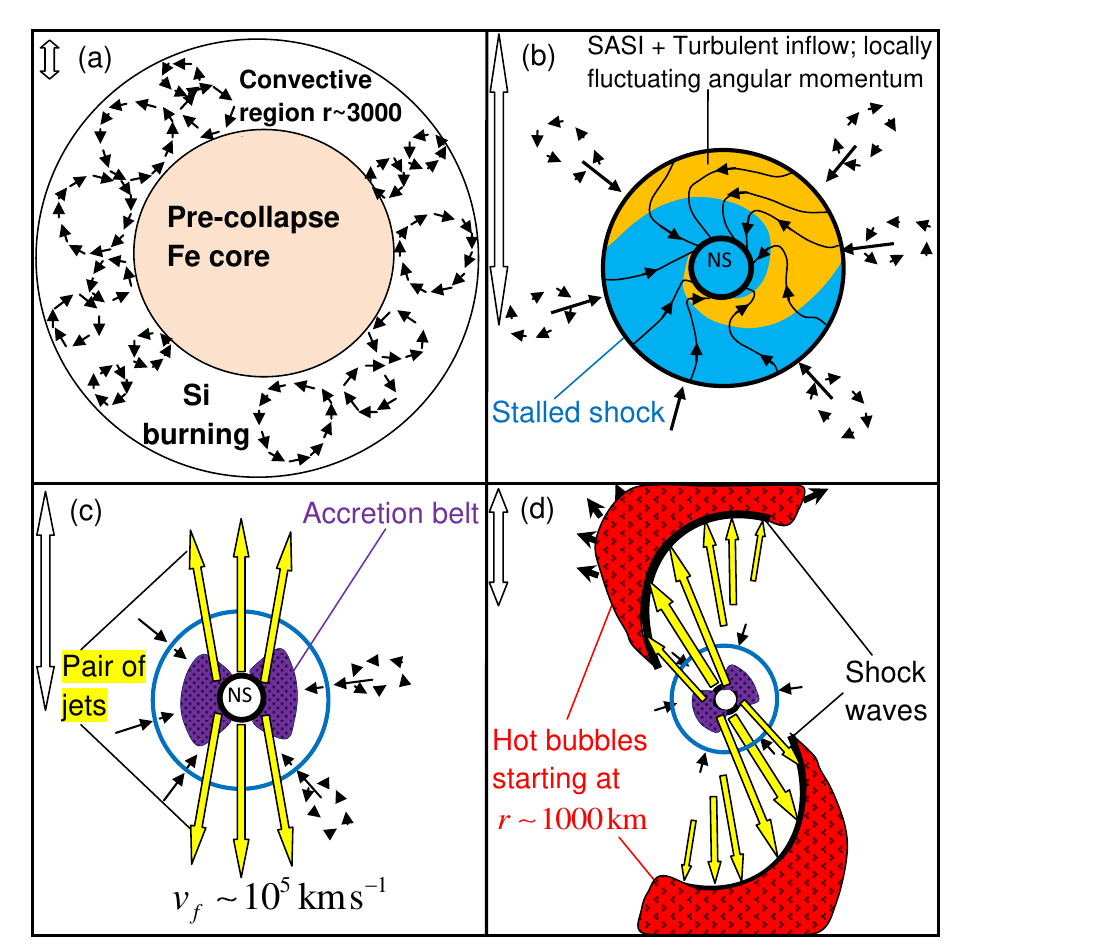}
\caption{A schematic presentation of the jittering jets explosion mechanism (From \citealt{GilkisSoker2015}). The two-sided arrow on the upper left of each panel represents a length of $\simeq 500 \km$. The four panels span an evolution time of several seconds. (a) The pre-collapse inner core just before collapse. The convective vortices in the silicon-burning shell (shown) and/or the oxygen-burning shell (not shown) of the pre-collapse core are the seeds of stochastic angular momentum fluctuations. (b) In-falling gas onto the newly born NS passes through the stalled shock and the spiral modes of the standing accretion shock instability (SASI) amplify the seed angular momentum perturbations. (c) The stochastic angular momentum variations of the accreted mass leads to the formation of an intermittent accretion disk/belt around the NS that launches jittering jets.  (d) The jittering jets are shocked and inflate hot bubbles that explode the core and then star. The entire jet-activity phase lasts for one to few seconds, and there might $\approx {\rm few} - 30$ jet-launching episodes. 
}
\label{fig:Jittering}
\end{figure}

In the \textit{jittering jets explosion mechanism} jets that the newly born NS (or BH) launches in varying directions in a stochastic manner explode the star (\citealt{Soker2010, PapishSoker2011}; for some later studies see, e.g., \citealt{PapishSoker2014Planar, GilkisSoker2015, Quataertetal2019, Soker2019SASI, Soker2020RAA, AntoniQuataert2022, ShishkinSoker2022, Soker2022SNR0540, Soker2022Boosting}). The NS launches these jets as it accretes mass via an accretion disk (or belt) with stochastic angular momentum variations. The source of these angular momentum variations is the pre-collapse stochastic convection motion in the core. The perturbations are amplified by instabilities above the newly born NS.
I schematically present the mechanism in Fig. \ref{fig:Jittering} (from \citealt{GilkisSoker2015}) and turn now to further discuss its properties (section \ref{subsec:AngularMomentumSource}).  

\subsection{The stochastic angular momentum source} 
\label{subsec:AngularMomentumSource}

The jittering jets explosion mechanism is an alternative to the much older delayed neutrino explosion mechanism that \cite{BetheWilson1985} suggested four decades ago to explain CCSNe. Hundreds of papers studied the neutrino-driven mechanism over the years (e.g.,  \citealt{Hegeretal2003, Janka2012, Nordhausetal2012, CouchOtt2013, Bruennetal2016, Jankaetal2016R, OConnorCouch2018, Mulleretal2019Jittering, BurrowsVartanyan2021, Fujibayashietal2021, Bocciolietal2022, Nakamuraetal2022}). 
Despite the very sophisticated three-dimensional (3D) hydrodynamical simulations of the delayed neutrino mechanism it still encounters some problems (which I will not elaborate on in this study; see, e.g., \citealt{Kushnir2015, Papishetal2015}). One of the main limitations of the delayed neutrino mechanism is that even when the explosion energy is scaled to observed CCSNe this mechanism is limited to explosion energies (mainly the kinetic energy of the ejecta) of $E_{\rm exp} < 3 \times 10^{51} \erg$ (e.g., \citealt{Fryer2006, Fryeretal2012, Papishetal2015, Ertletal2016, Sukhboldetal2016, Gogilashvilietal2021}). 

The fundamental ingredients and outcomes of the jittering jets explosion mechanism are as follows. 
\begin{enumerate}
\item  \textit{Source of angular momentum.} The source of angular momentum is the stochastic convective motion in the precollapse core (e.g., \citealt{GilkisSoker2014, GilkisSoker2016, ShishkinSoker2021}) or envelope (e.g., \citealt{Quataertetal2019}). These seed perturbations are amplified by instabilities behind the stalled shock inside a radius of $r \simeq 100 \km$ (\citealt{Soker2019SASI, Soker2019JitSim}), mainly by the spiral standing accretion shock instability (spiral SASI, e.g.,  \citealt{Andresenetal2019, Walketal2020, Nagakuraetal2021, Shibagakietal2021}, for some simulations of the spiral SASI).
The large stochastic angular momentum fluctuations of the mass that the newly born NS accretes allow the formation of intermittent accretion disks that launch the jittering jets (e.g., \citealt{PapishSoker2011,  GilkisSoker2014, GilkisSoker2015, Quataertetal2019}; accretion belts are also possible, \citealt{SchreierSoker2016}).
\item \textit{The crucial role of magnetic fields.} Magnetic fields are essential ingredients in the launching of the jets, and must be included in future numerical simulations of the jittering jets explosion mechanism (e.g., \citealt{Soker2018arXiv, Soker2019SASI, Soker2020RAA}). 

\item \textit{Coupling to neutrino heating.} Although neutrino heating does not play the main role in powering CCSNe according to the jittering jets explosion mechanism, neutrino heating boosts the energy of the jittering jets \citep{Soker2022Boosting}. 
\item \textit{No failed CCSNe.} As core and envelope convection zones exist in \textit{all CCSN progenitors} (e.g., \citealt{ShishkinSoker2021, AntoniQuataert2022, ShishkinSoker2022}), according to the jittering jets explosion mechanism there are no failed CCSNe (e.g., \citealt{Soker2017TwoPromissing}). The observational finding by \cite{ByrneFraser2022} of no failed CCSNe indirectly supports the jittering jets explosion mechanism. 
\item \textit{Feedback mechanism.} The jittering jets explosion mechanism operates through a negative feedback mechanism (see review by \citealt{Soker2016Rev}). In the negative jet feedback mechanism the jets regulate their power by their effect on the mass accretion rate onto the compact object that launches the jets. Specifically, if the mass accretion rate increases so is the power of the jets. However, the interaction of the jets with the ambient gas, which is the reservoir of the accreted mass, expel and inflate the ambient gas. This in turn reduces the mass accretion rate, and hence the jets' power, closing the negative feedback cycle. As long as there is no explosion, falling material feeds the accretion disk that launches the jets. Because there is no hundred percent conversion efficiency of jets' energy to unbind the ejecta, i.e., the ejecta expands at velocities much above the binding energy, this explains why typical CCSN explosion energy is about several times the ejecta binding energy. 
\item \textit{Typical properties of the jittering jets.}
During an explosion by jittering jets there might $\approx {\rm few} - 30$ jet-launching episodes.  Some characteristic values for most, but not all, CCSNe are as follows \citep{PapishSoker2014a}. Jets are launched with velocities of $\simeq 10^5 \km \s^{-1}$ (neutrino observations limit the jets in most cases to be non-relativistic, e.g. \citealt{Guettaetal2020}). 
In total the jets carry an energy of $\simeq 10^{51} \erg$. The explosion time might be $\simeq 1 - {\rm few} \s$. Each individual jet-launching episode carries a mass of $\approx 10^{-3} M_\odot$ and lasts for a time period of $\simeq 0.01-0.1 \sec$. Each accretion disk of an episode has a mass of $\approx 10^{-2} M_\odot$. During the entire jet-driven explosion process the newly born NS accretes a mass of $\approx 0.1 M_\odot$ through intermittent accretion disks. 
From one episode to the next the jets might change axis direction by a very large angle. 
The number of episodes and the changes in jets' directions depend on the properties of the convection motion, the binding energy of the ejecta, and the precollapse core rotation. 
\item \textit{Smooth connection to superluminous supernovae (SLSNe).}
When the precollapse core rotation is fast enough to allow a stable accretion disk to form there will be only one axis (with small jittering around this axis) along which the NS launches jets. The jets will be inefficient in removing core and envelope mass from the equatorial plane. As a result of that the NS accretes more mass, and might become a BH. Therefore, not only that according to the jittering jets explosion mechanism (more generally, according to the jet feedback explosion mechanism) there are no failed CCSNe, but rather the formation of a BH implies a very energetic CCSN (e.g., \citealt{Gilkisetal2016Super}). The outcomes are luminous CCSNe (LSNe) or superluminous CCSNe (SLSNe) that have bipolar morphologies. I study these CCSNe in section \ref{sec:CCSNeFixed}. Not only the jet-driven explosion mechanism connects regular CCSNe to LSNe and SLSNe, but it also connects the fixed-axis (with small jittering) jets model to CEJSNe that I discuss in section \ref{sec:CEJSNe}. 
\item \textit{Signatures of jets in supernova remnants (SNRs).}
Many SNRs possess signatures of jets, as is the expectation in the jittering jets explosion mechanism. These features include `Ears' that carry a small fraction of the total explosion energy and SNRs with point-symmetry morphologies. I devote section \ref{subsec:SNRSignaturesJets} to review these signatures.
\item \textit{Natal kick of the neutron star}. \cite{BearSoker2018kick} adopted the tug-boat mechanism where a massive clump that the explosion process ejects from the core gravitationally pulls and accelerates the NS 
(e.g., \citealt{Nordhausetal2010, Wongwathanaratetal2013, Janka2017}; different explanations to natal kick velocity exist, e.g., \citealt{Yaoetal2021} and \citealt{Xuetal2022} for recent studies). \cite{BearSoker2018kick} list two processes to explain why in the jittering jets explosion mechanism the natal kick velocity avoids small angles to the jets' axis. ($i$) The jets prevent the formation of dense clumps along their propagation direction; ($ii$) Dense clumps supply the gas to the accretion disk that launches the jets and therefore concentrate in a plane perpendicular to the jets. One or more of the dense clumps are ejected and pull the NS. In Fig. \ref{fig:KickDistribution} I present the cumulative distribution function ${\rm W}_{\alpha}$ of the projected angles between the NS natal kick direction and the jets'-axis, and compare with the expectation of random distribution and a distribution of perpendicular angles only (from \citealt{Soker2022SNR0540}). Clearly the kick velocity avoids small angles to the jets' axis. 
\begin{figure}[!t]
\includegraphics[trim=3.6cm 8.2cm 0.0cm 8.0cm ,clip, scale=0.63]{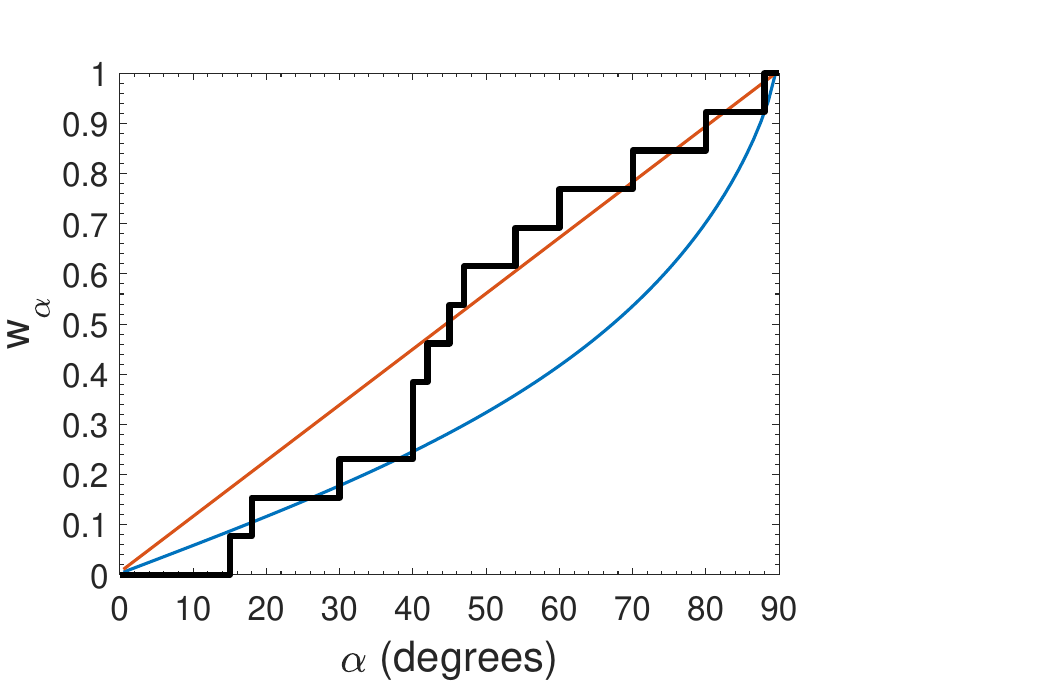}
\caption{The cumulative distribution function ${\rm W}_{\alpha}$ of projected angles between the NS natal kick direction and the jets'-axis for 13 SNRs (from \citealt{Soker2022SNR0540}).
The lower blue line represents the cumulative distribution function where in all SNRs the three-dimensional NS kick velocity is perpendicular to the jets' axis.
The straight orange line represents the random cumulative distribution function.
}
\label{fig:KickDistribution}
\end{figure}
\end{enumerate}

The above properties distinguish the jittering jets explosion mechanism from other jet-driven explosion mechanisms (e.g., \citealt{LeBlancWilson1970, Khokhlovetal1999, Aloyetal2000, MacFadyenetal2001, Burrowsetal2007, BarkovKomissarov2008, Wheeleretal2008, Maedaetal2012, LopezCamaraetal2013, BrombergTchekhovskoy2016,  Nishimuraetal2017, Sobacchietal2017, Grimmettetal2021, Gottliebetal2022a, Gottliebetal2022Wobbling, Perleyetal2022, Ghodlaetal2022}). In particular, these mechanisms assume a rapidly rotating precollapse core, something that is not required in the jittering jets explosion mechanism.  

\cite{Gottliebetal2022a} conduct magnetohydrodynamic simulations of a collapsing core onto a central rapidly rotating BH of $4 M_\odot$. Their study is relevant to points 1-4 and 7 above. \cite{Gottliebetal2022a} find that jet launching requires both rapid rotation and strong magnetic fields. In cases of too low angular momentum only weak jets are formed or not at all. In the jittering jets explosion mechanism most CCSNe have a central NS and therefore the accretion process is different than that onto a rapidly rotating BH. In addition, neutrino heating, which \cite{Gottliebetal2022a} do not include in their simulations, boosts the jets \citep{Soker2022Boosting}. Therefore I expect that a NS can launch jets event in cases of angular momentum that is somewhat below the minimum to form a thin accretion disk, i.e., it forms an accretion belt \citep{SchreierSoker2016}). The formation of a BH in the jittering jets explosion mechanism requires rapidly rotating pre-collapse core (point 7 above), and therefore it is along the results of \cite{Gottliebetal2022a}. I expect strong jets (although not necessarily that break out to give a gamma ray burst).  

Another possible effect is the formation of double-peak lightcurves in cases where strong jets transport some of the newly synthesised $^{56}$Ni to the outer regions of the ejecta \citep{OrellanaBersten2022}. \citep{OrellanaBersten2022} demonstrate how the jet-delivered $^{56}$Ni in the outer regions of the ejecta can power the first peak, before the inner $^{56}$Ni powers the second (regular) peak in of lightcurve. It is not clear whether individual jets of the jittering jets are sufficiently strong to explain this effect, or whether only stronger jets that also explode bipolar CCSNe can power this process. 

Before closing this subsection I point out that at present the absence of self-consistent magnetohydrodynamical simulations that show that the jittering jets explosion mechanism actually works is still a drawback of the model. Future simulations that will require huge computer resources are still needed to establish this explosion mechanism. 
  
\textbf{$\bigstar$ Summary of section \ref{subsec:AngularMomentumSource}.}  The development of the jittering jets explosion mechanism brought the call to change paradigm from neutrino-driven explosions to jet-driven explosions of all CCSNe (e.g., \citealt{Papishetal2015, Bearetal2017}). (\citealt{Izzoetal2019} and \citealt{Piranetal2019} adopted this earlier call but with a weaker emphasize of the jets' major role.)
In the present review I strengthen this call. 

\subsection{Signatures of jittering jets in SNRs} 
\label{subsec:SNRSignaturesJets}

\subsubsection{Large scale structures} 
\label{subsubsec:LargeScale}

Many CCSN remnants possess morphological features that are imprints of jets (e.g., \citealt{Bearetal2017, GrichenerSoker2017, YuFang2018, Luetal2021, Soker2022SNR0540}). One of the clearest example is the north-east jet of Cassiopeia A (e.g., \citealt{Grefenstetteetal2017}). 
Some examples of other claims for signatures of jets in SNRs include 
Vela SNR (e.g., \citealt{Garciaetal2017, Sapienzaetal2021}) and IC 443 (e.g., \citealt{Grecoetal2018}). 
I note that the jets' axis that \cite{Sapienzaetal2021} and  \cite{Garciaetal2017} identify in the Vela SNR is not the one that \cite{GrichenerSoker2017} take to be the jets' axis. However, according to the jittering jets explosion mechanism there are many more than one jets' axis, and the question is how many of these reveal themselves in the SNR. 
   
I concentrate here on similarities of SNR morphologies with morphologies of planetary nebulae (PNe; for a review on the shaping of PNe see, e.g., \citealt{DeMarco2009}). This approach allows the usage of morphologies of PNe that are thought to result from jets to argue for jets in CCSNe (e.g., \citealt{BearSoker2017, Bearetal2017, Akashietal2018, BearSoker2018SN1987A, Soker2022SNR0540}). Although other models exists for the formation of ears that do not include jets I here adopt the view that ears are shaped by jets and present arguments to support this. The other model include the simulations of \cite{Blondinetal1996} of a spherical CCSN ejecta that runs into a CSM with a density gradient, and a spherical explosion into a CSM with a dense ring (e.g.,  \citealt{Chiotellisetal2021, Ustamujicetal2021}). In the later model the ears are actually a ring protruding from the main ejecta, rather than two opposite polar ears.    

\begin{figure*}
\centering
\includegraphics[trim= 0.cm 0.cm 0.0cm 0.9cm,scale=0.64]{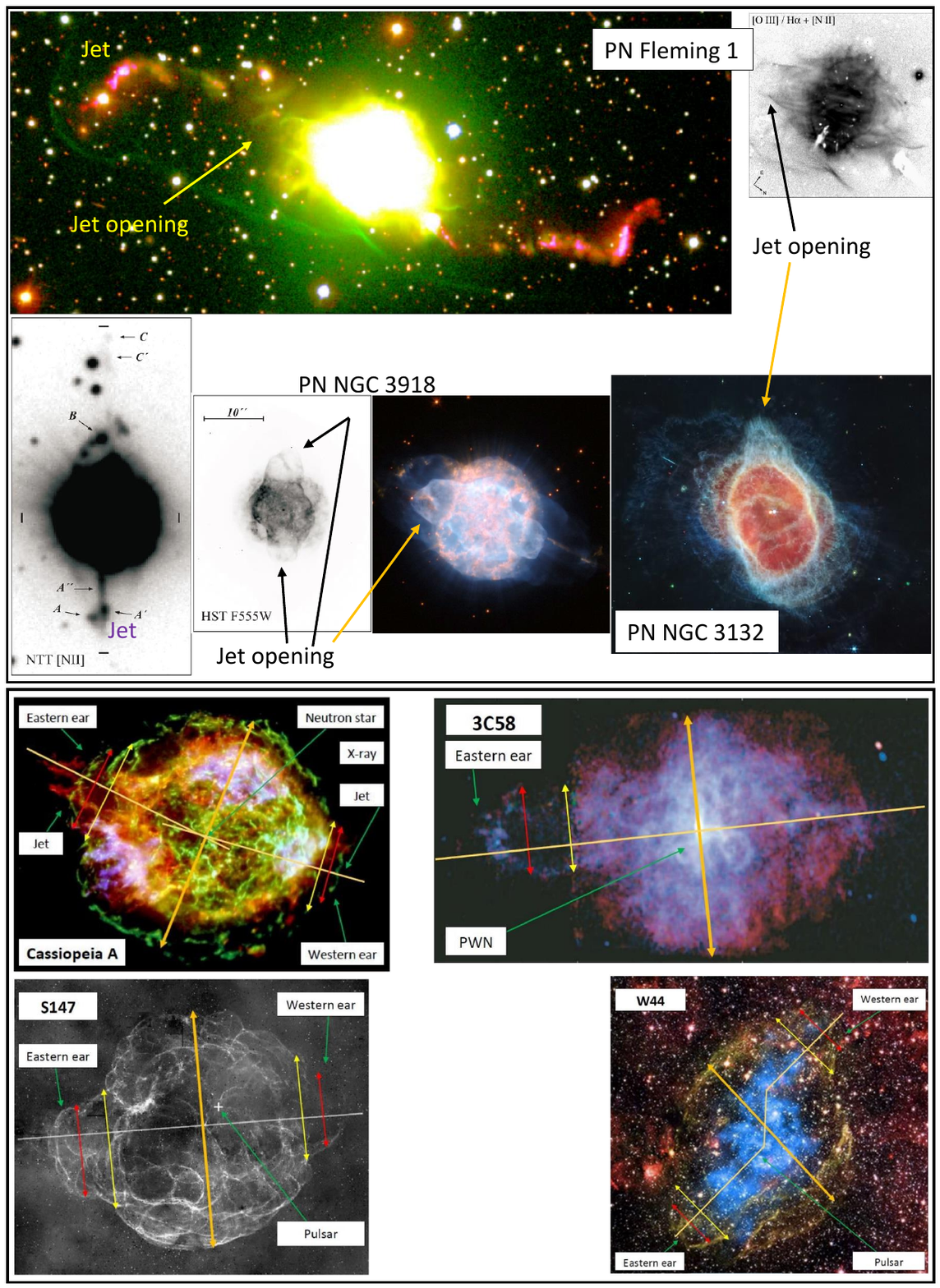}
\caption{Comparing some signatures of jets in planetary nebulae (PNe) and CCSNe with emphasize on `Ears', which in some cases can be identified as jet opening as I mark in the upper rows. 
Upper row: Images of the PN Fleming 1 from \cite{Boffinetal2012}. 
Second row: Left two panels are images of PN NGC 3918 from \cite{Corradietal1999}, and the third panel from left is an HST image of NGC 3918 (ESA/Hubble and NASA; not to scale with the left images). Right panel is a Webb’s MIRI image in the mid-infrared (NASA, ESA, CSA, and STScI). 
The two lower rows are of SNRs with marks from \cite{GrichenerSoker2017} that define and emphasize the ears. These marks were used to estimate the energy of the jets that inflated the ears. 
Third row left: An X-ray image taken from the Chandra gallery (based on \citealt{Hwangetal2004}). Red, blue and green represent Si He$\alpha$ (1.78-2.0 keV), Fe K (6.52-6.95 keV), and  4.2-6.4 keV continuum, respectively. 
Third row right: ACIS/Chandra image of SNR 3C58 in the energy bands $0.5 - 1.0 \keV$ (red), $1.0-1.5 \keV$ (green), and $1.5-10 \keV$ (blue), based on \cite{Slaneetal2004}; 
Lower row left: An H$\alpha$ image of the SNR Semeis~147 taken from \cite{Gvaramadze2006} who reproduced an image from \cite{Drewetal2005}.
Lower row right:  Composite image of the SNR W44 taken from the Chandra gallery. The cyan represents X-ray (based on \citealt{Sheltonetal2004}), while the red, blue and green represent infra-red (based on NASA/JPL-Caltech). The three beige thick lines  schematically define the S-shape of this SNR that hint at jet precession.  
}
 \label{Fig:JetsPNeEars}
 \end{figure*}

In Fig. \ref{Fig:JetsPNeEars} I compare some SNRs to some PNe. I present four SNRs with the marks of the `Ears'. Ears are two opposite protrusions from the main part of the SNR of from the PN main shell that are smaller than the main nebula and with a cross section that decreases monotonically from the base of an ear at the shell to its far end (e.g., \citealt{AkashiSoker2021Ears}; small departures from this definition are possible.)  The upper two rows of Fig. \ref{Fig:JetsPNeEars} present three PNe, two of which show jets. The region where a jet breaks out from the main shell has a shape that I term  `jet opening', which in most cases obeys the definition of an ear. \cite{GrichenerSoker2017} argue that the ears in the four SNRs in the two lower rows, as well as other SNRs that they study, are actually jet openings. Namely, they were shaped by jets (also \citealt{Bearetal2017}). They further estimate that about third of all CCSN remnants (CCSNRs) have clear ears. 

\begin{figure*}
 \centering
\includegraphics[trim= 0.6cm 3.0cm 0.0cm 2.0cm,scale=0.90]{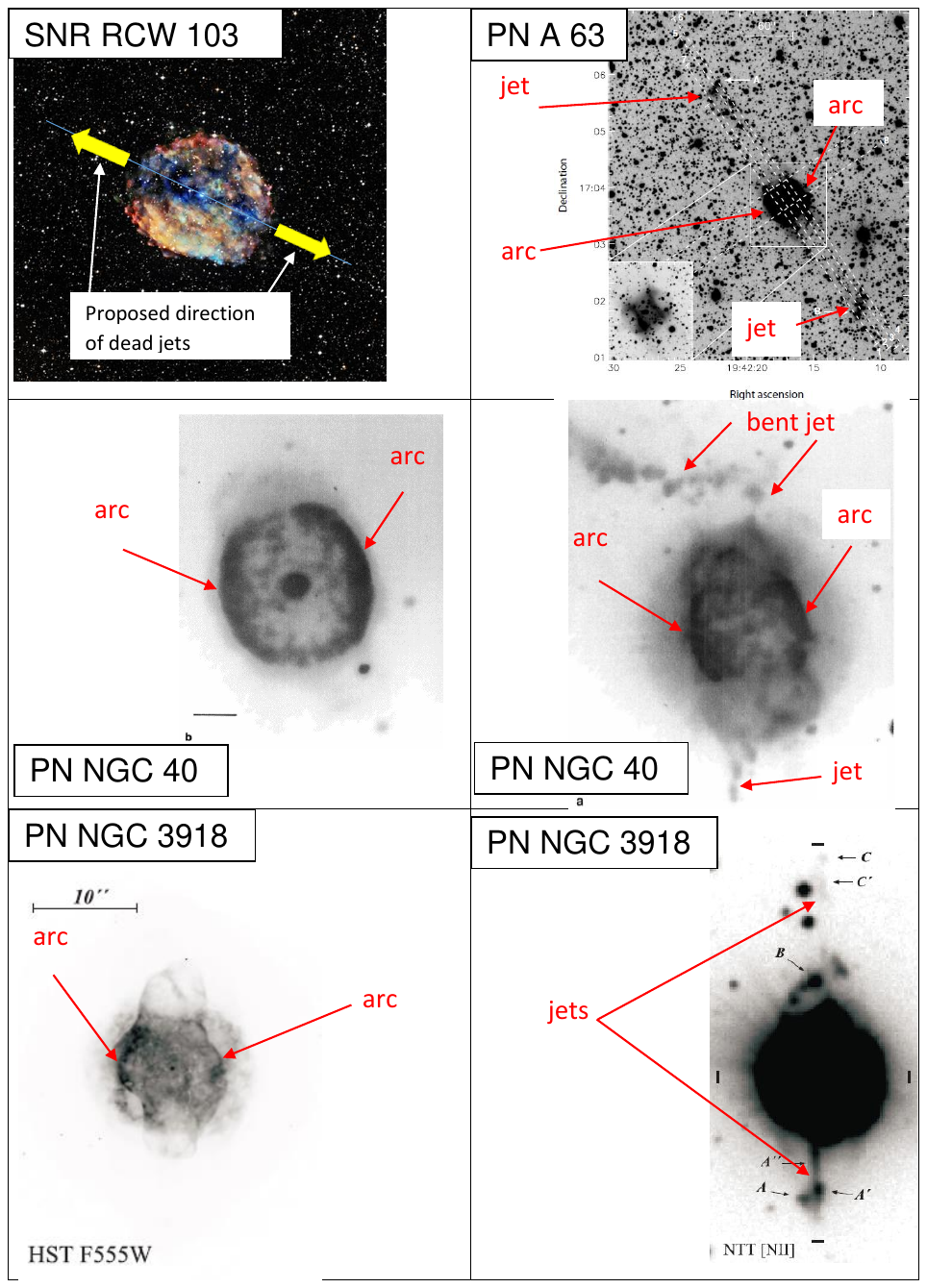}
\caption{Comparing some signatures of jets in the CCSN remnant (CCSNR) RCW~103 and three PNe emphasizing two opposite arcs that are the projection of an axially-symmetric barrel-shaped nebula. This figure is from \cite{Bearetal2017}. 
The composite image of SNR RCW~103 is of X-ray (colors according to energy bands: low=red, medium=green, highest=blue) together with an optical image from the Digitized Sky Survey (image from the Chandra website; based on \citealt{Reaetal2016}). 
The proposed original directions of the, already dead, jets in the CCSNR RCW~103 are marked by yellow thick arrows.
The three PNe are A~63 (image from \citealt{Mitchelletal2007}), NGC~40 (images from \citealt{Meaburnetal1996}), and NGC~3918 (images from \citealt{Corradietal1999}; images are rotated).  
}
 \label{Fig:JetsPNeArcs}
 \end{figure*}

Fig. \ref{Fig:JetsPNeArcs} that I take from \cite{Bearetal2017} emphasises the barrel shape, i.e., a hollow cylinder with convex-shaped surface. The three PNe in that figure have clear indications that jets shaped the barrel-shaped main nebula. \cite{Bearetal2017} suggest that this was the case in the SNR RCW 103 as well. \cite{Akashietal2018} simulated this shaping by jets. 

According to the jittering jets feedback mechanism the jets collide with the core, explode the core, and by that explode the entire star. The explosion from that time on is actually similar to that expected in the delayed neutrino mechanism. The negative feedback cycle implies that when the entire core is ejected the high-mass-accretion rate onto the newly born NS ceases, although late fallback with low-mass-accretion rate might continue (see section \ref{sec:PostExplosion}). However, at the time of core explosion there is mass that already flows towards the newly born NS. As a result of that one or two jet-launching episodes might occur after core explosion. These jets expand to large distances before they encounter the already diluted ejecta. Therefore, these last jets might penetrate into the ejecta and even go out from the main ejecta. As such, these jets might leave imprints on the ejecta, such as ears and hollowed barrel-shaped SNR.

\begin{figure}
 \centering
\includegraphics[trim= 1.7cm 1.1cm 0.0cm 0.0cm,scale=0.20]{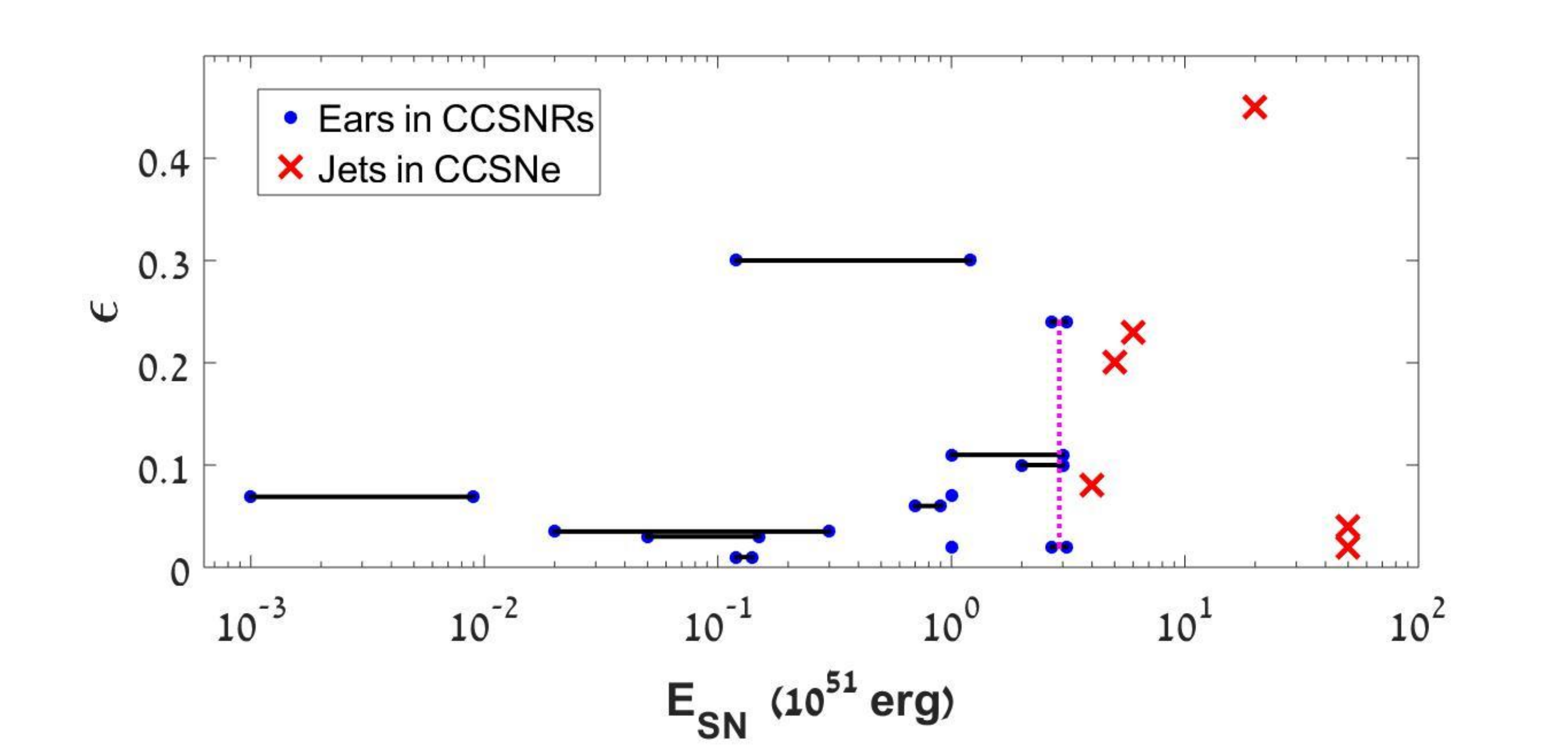}
\caption{The energy to inflate the ears in 11 CCSNRs and the jets' energy in 6 SLSNe relative to the explosion energy, $\epsilon_{\rm ears}$ (blue dots) and $\epsilon_{\rm SN, jets}$ (red crosses), respectively, as function of the explosion energy $E_{\rm SN}$. The black horizontal lines connect dots for the same SNR of the minimum and maximum values of the explosion energy.  The vertical dotted pink line connects the two options for the western ear of N49B. The red crosses are for the ratio $\epsilon_{\rm SN,jets} \equiv E_{\rm SN,jets}/E_{\rm SN}$, where the values of $E_{\rm SN,jets}$ and $E_{\rm SN}$ are taken from Table 1 of \cite{Piranetal2019}. 
The graph is from \cite{Bearetal2017} who give all details. 
}
 \label{Fig:JetsEnergyGraph}
 \end{figure}

From an energetic point of view, the two jets of each jet-launching episode carry only a few per cents to few tens of per cents of the explosion energy (point 6 in section \ref{subsec:AngularMomentumSource}). Therefore, the energy that inflate the ears has that typical value. \cite{GrichenerSoker2017} built a simple geometrical scheme (for that purpose are the marks on the lower panels in Fig. \ref{Fig:JetsPNeEars}) to estimate the energy of the two jets (combined) that inflated the ears relative to that of the CCSN explosion, $\epsilon_{\rm ears}$. \cite{Bearetal2017} present these fractional energies against the explosion energy. They also added to that graph the relative jets' energy in six SLSNe that \cite{Piranetal2019} study. I present their graph in Fig. \ref{Fig:JetsEnergyGraph}.
\cite{Bearetal2017} finding that the energy of the jets that inflated the ears is only a small fraction of the explosion energy is compatible with the expectation of the jittering jets explosion mechanism.
In section \ref{sec:CCSNeFixed} I discuss jets that have a constant axis and therefore shape the ejecta to become bipolar, namely with two very large inflated bubbles rather than ears. 

\textbf{$\bigstar$ Summary of section \ref{subsubsec:LargeScale}.}
About third of all CCSNRs have ears. Comparison to PNe (Figs. \ref{Fig:JetsPNeEars} and \ref{Fig:JetsPNeArcs}) suggests that the ears were shaped by jets. These jets most likely played a significant role in the explosion process. The energy to inflate the ears is $\simeq 1-10\%$ of the explosion energy  (Fig. \ref{Fig:JetsEnergyGraph}), compatible with the expectation of the jittering jets explosion mechanism.  

\subsubsection{Clumps and filaments} 
\label{subsubsec:ClumpsFilaments}

CCSNR structures are inhomogeneous, e.g., in having filaments, arcs, clumps, and as discussed above some have ears. In many cases different elements are concentrated in different zones. Examples of CCSNRs with filaments and clumps include Cassiopeia A (e.g., images by \citealt{Grefenstetteetal2017, Leeetal2017}) SNR~G$292.0+1.8$ (e.g., \citealt{Parketal2002, Parketal2007}), Vela (e.g.,  \citealt{Aschenbachetal1995, Garciaetal2017}), and SNR~W49B (e.g., \citealt{Lopezetal2013, Sanoetal2021}). 
The case of SNR~W49B is interesting. \cite{GonzalezCasanovaetal2014} performed hydrodynamical simulations and argued that a jet-driven CCSN explosion can explain the distribution of metals, like silicon and iron, in SNR~W49B. The jets' axis they propose (also  \citealt{Micelietal2008, Lopezetal2011, Lopezetal2013}) and the jets' axis that \cite{BearSoker2017} suggest to explain the morphology SNR~W49B are perpendicular to each other. Future studies should settled this disagreement. 

According to the delayed neutrino explosion mechanism  instabilities that develop during the explosion process alone form clumps and filaments in the inner ejecta (e.g., \citealt{Jankaetal2017, Wongwathanarat2017, Gableretal2021, Sandovaletal2021, Larssonetal2021}). 
According to the jittering jets explosion mechanism the shaping is due both to instabilities and to jets (e.g., \citealt{Soker2022SNR0540}). 

This dispute between the two explosion mechanism supporters is best demonstrated in SN~1987A that has a clumpy and filamentary ejecta as recent observations show (e.g., \citealt{Franssonetal2015, Franssonetal2016, Larssonetal2016, Abellanetal2017, Matsuuraetal2017}).
\cite{Kjaeretal2010} argued for shaping by instabilities alone. However, \cite{Soker2017TwoPromissing} and later \cite{Abellanetal2017} found that neutrino driven explosion simulations (e.g., \citealt{Wongwathanaratetal2015}) do not fit all observations of SN~1987A. 
\cite{BearSoker2018SN1987A} studied some morphological features of SN~1987A \citep{Abellanetal2017} alongside morphologies of some other CCSNRs and of PNe. They strengthened earlier claims that  jittering jets likely played a crucial role in the explosion and shaping of SN~1987A.  \cite{BearSoker2018SN1987A} noted that the structure of the ejecta of SN~1987A from \cite{Abellanetal2017} rule-out the old claim of \cite{Wangetal2002} for two opposite non-jittering jets that exploded SN~1987a. More recently, \cite{Onoetal2020} and \cite{Orlandoetal2020} performed 3D hydrodynamical simulations of evolution of SN~1987A and concluded that jet-driven explosion with the jets' axis in the plane of the inner CSM ring best reproduce the explosion morphology and element distribution. \cite{BearSoker2018SN1987A} took the jets' axis to be at an angle to the plane of the inner ring. This disagreement on the jets' axis should be settled by further exploration of SN 1987A. 

The dispute exists also in the analysis of SNR Cassiopeia~A, for which \cite{Wongwathanaratetal2015}, \cite{Orlandoetal2021} and \cite{Orlandoetal2022} argued that the neutrino-driven explosion mechanism can account for the distribution of some metals, while in \cite{Soker2017TwoPromissing} I argued that jets seem to have played a crucial role is the shaping of Cassiopeia~A during its explosion. \cite{Orlandoetal2016} already argued that instabilities alone cannot account for all morphological features of Cassiopeia~A.

\begin{figure*}[ht]
 \centering
\includegraphics[trim= 0.3cm 7.0cm 0.0cm 3.0cm,scale=0.80]{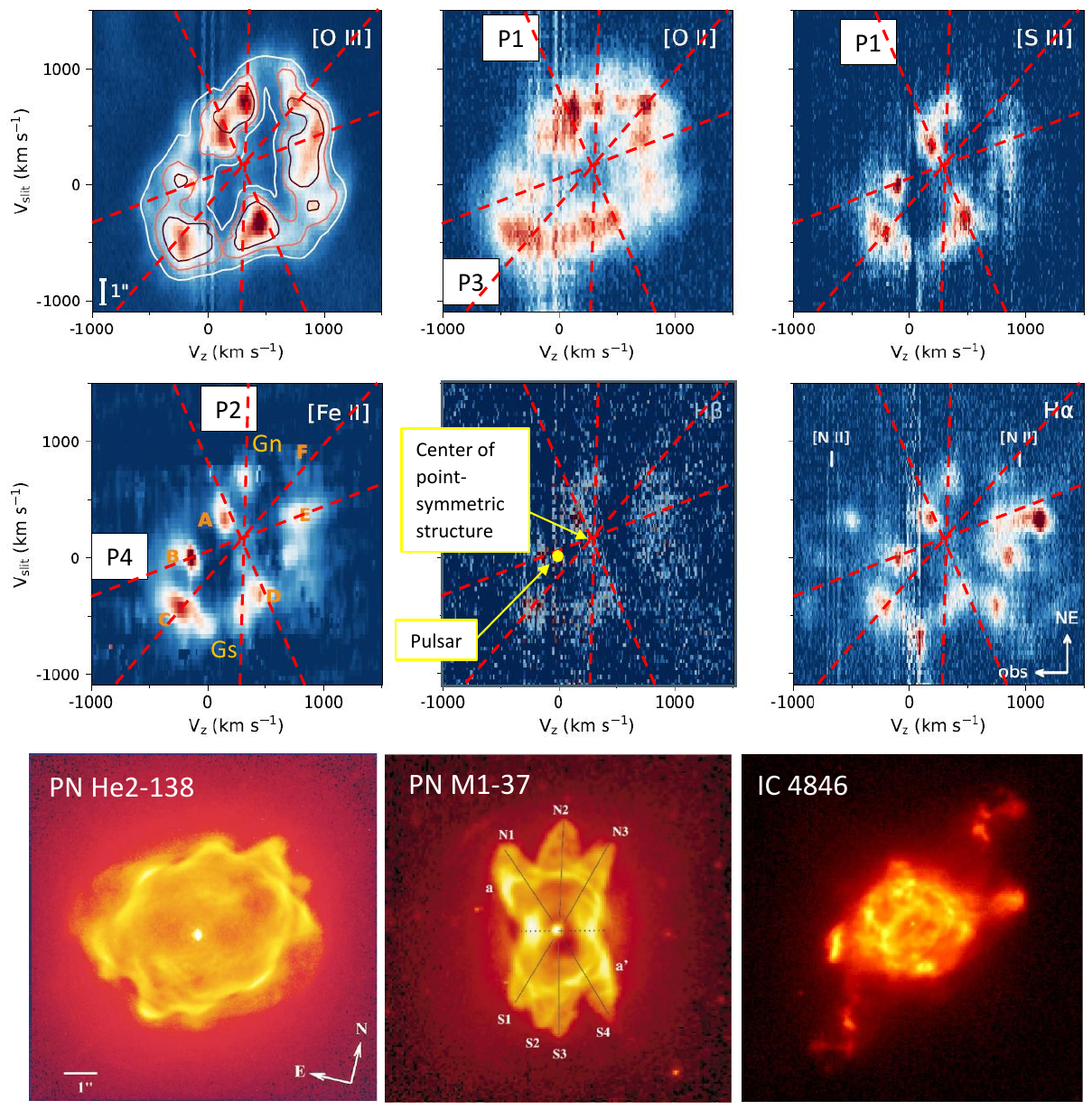}
\vskip +0.500 cm
\caption{Upper two rows: Two-dimensional velocity maps of SNR~0540-69.3 taken from figure 4 of \cite{Larssonetal2021}.
$v_{\rm z}$ is the velocity along the line of sight, while $v_{\rm slit}$ is the velocity along the slit (see their paper for details). In \cite{Soker2022SNR0540} I added the same four lines, P1 to P4, on all panels to mark four pairs of opposite clumps that together form a point-symmetric morphology (A-D; B-E; C-F; Gn-Gs). In the lower-left panel I added the marks to clumps Gn and Gs, while clumps A to F are from \cite{Larssonetal2021}. The pulsar is at $v_{\rm slit}=0$ in these panels. 
Lower row: Three PNe, PN He2-138 (PN~G320.1-09.6), PN~M1-37 (PN~G002.6-03.4) and IC 4846 (PN~G027.6-09.6), that have point-symmetric morphologies that  \cite{SahaiTrauger1998}, \cite{Sahai2000} and \cite{Sahaietal2011}, respectively, attribute to jets. The three lines on the M1-37 image are from  the original image of \cite{Sahai2000}.
For more details see \cite{Soker2022SNR0540}.
}
 \label{Fig:PointSymmetry}
 \end{figure*}

Most recently I \citep{Soker2022SNR0540} analyzed the structure of SNR~0540-69.3 from the observations of \cite{Parketal2010} and \cite{Larssonetal2021}. Although \cite{Larssonetal2021} argued that instabilities alone can account for the ejecta structure, I argued for jittering jets (in addition to instabilities). 
I identified a point-symmetric morphology in the VLT/MUSE velocity map in a plane along the line of sight (perpendicular to the plane of the sky), as I show in the upper two rows of Fig. \ref{Fig:PointSymmetry} that are based on the results of \cite{Larssonetal2021}.
Comparing the four pairs of two opposite clumps that the images in Fig. \ref{Fig:PointSymmetry} show to point-symmetric PNe, three of which I present in the lower row of Fig. \ref{Fig:PointSymmetry}, 
brought me to propose that two or four pairs of jittering jets shaped the inner ejecta of SNR~0540-69.3. 
In \cite{Soker2022SNR0540} I further argued that both jets and instabilities mix elements in the ejecta of CCSNe. 

Figs \ref{Fig:JetsPNeEars} and \ref{Fig:PointSymmetry} show that jets change directions in PNe. Here I concentrate on the jittering of jets in CCSNe that is caused by the convective motion of the precollapse core. There are other effects than can change the jets' axis not only in CCSNe, but also in a variety of systems from binary systems, like progenitors of PNe, and up to jets of active galactic nuclei. In binary systems the gravitational force of the companion can cause the disk to precess. However, I would like to comment on an effect where the jets have feedback on the angular momentum. If the interaction of the jets with the ambient medium causes mater to fall towards the accretion disk this material will have a velocity with a small angle to the original jets' axis. As a result of that the angular momentum of this gas is at a large angle to the angular momentum of the accretion disk that launches the jets. As this falling gas feeds the accretion disk it changes its angular momentum axis, hence the jets' axis. In cooling flow clusters of galaxies active galactic nucleus jets might cause dense clumps to feed the accretion disk around the supermassive BH \citep{Soker2018JitteringCF}. I termed this jittering jets by negative angular momentum feedback \citep{Soker2021JitteringCF}.
\cite{Gottliebetal2022Wobbling} conduct 3D magneto-hydrodynamical simulation of a CCSN and find that bound gas from the two jet-inflated cocoons feeds the accretion disk and tilts it somewhat. This lead to jets with relatively small jittering that they term wobbling jets. 

\textbf{$\bigstar$ Summary of section \ref{subsubsec:ClumpsFilaments}.} Both jets and instabilities shape clumps and filaments in the ejecta of CCSNe. Instabilities alone cannot account for some properties, like point-symmetric morphologies and some elements distributions. In most cases jets jitter by moderate to large angles between different jet-launching episodes.   

\section{Bipolar core collapse supernovae} 
\label{sec:CCSNeFixed}

\subsection{Super-luminous supernovae (SLSNe)} 
\label{subsec:SuperluminousSNe}

Two sources feed angular momentum to the mass that the NS accretes, the precollapse core rotation (point 7 in section \ref{subsec:AngularMomentumSource}) and the angular momentum fluctuations of the precollapse core convection that are amplified by instabilities (point 1 of section \ref{subsec:AngularMomentumSource}).  When the precollapse specific angular momentum is large enough to allow the formation of a stable accretion disk around the newly born NS the jittering is relatively small and occurs around a fix axis along the precollapse angular momentum of the core. This is the fixed-axis jets explosion process. Because all jets share more or less the same axis they very efficiently expel gas from the polar directions surroundings, but not from the equatorial plane surroundings. The outcome is a bipolar explosion, i.e., the morphology is of two opposite large bubbles (rather than ears) with a waist between them.

Because of the low efficiency in mass ejection the total energy that the jets carry is much larger than the binding energy of the ejected gas. Therefore, the fixed-axis jets explosion process leads to a super-energetic CCSN. If even a small fraction of the kinetic energy is channelled to radiation the outcome is a LSNe or a SLSN. The prediction of the jittering jets explosion mechanism, or more generally the jet-feedback explosion mechanism, is that most SLSNe and a large fraction of LSNe are bipolar. I adopt here the definition (e.g., \citealt{Gomezetal2022}) that SLSNe have peak r-band magnitude of $M_{\rm r} < -20$, and LSNe have peak r-band magnitude of $M_{\rm r} =-19$ to $-20$.

The accretion of equatorial mass can lead to the formation of a BH. Therefore, according to the jittering jets explosion mechanism the formation of BHs in CCSNe comes along with super-energetic CCSNe, rather than with failed CCSNe (e.g., \citealt{Gilkisetal2016Super, Soker2017TwoPromissing}).

\cite{Chugaietal2005} suggested a bipolar nickel ejecta, i.e., two opposite jets of $^{56}Ni$, inside a spherical hydrogen-rich envelope for the type IIP SN 2004dj. They estimated the nickel mass to be $\approx 0.02 M_\odot$. \cite{UtrobinChugai2019} suggested a similar model for the type IIP SN 2016X but with a nickel mass of $\approx 0.03 M_\odot$. They took the explosion to be bipolar, but not to the degree of shaping the spherical hydrogen shell. These CCSNe are not LSNe/SLSNe. Because of the spherical envelope such a scenario might also take place with large jittering and pre-collapse core with moderate rotation velocity. Namely, the nickel bipolar structure is the last jet-launching episode, but a relatively massive one, in the jittering jets explosion mechanism. 

\cite{Hungerford2003} conducted 3D hydrodynamical simulation of jet-driven explosion and studied the effects of such asymmetrical explosions on mixing and gamma ray line emission. \cite{Hungerford2005} studied these effects in case of a ‘‘single-lobe’’ CCSN explosion as the delayed neutrino mechanism predicts in some case, and \cite{Wollaegeretal2017} examined the influence of single-lobe explosion on the emission from UV to IR. In that respect I note that two opposite jets must not always be equal, and even jittering jets can be unequal. The two opposite lobes of the SNR W50 that the jets from SS 433 inflate are highly unequal (e.g., \citealt{Dubneretal1998}), and so are the two unequal opposite lobes of the proto-planetary nebula OH 231.8+4.2  that were most likely shaped by jets (e.g., \citealt{Bujarrabaletal2002}).  

As with some other (but not all) issues, the jittering jets explosion mechanism and the delayed neutrino explosion mechanism depart also on their explanation of SLSNe. The jittering jets explosion mechanism attribute most of the energy of SLSNe to powering by jets, even when there are other processes, like ejecta-CSM collision (section \ref{sec:PostExplosion}) and a magnetar, which I discuss next.  
  
Most fittings of SLSNe lightcurves that assume the delayed neutrino explosion mechanism require extra energy sources because this mechanism cannot explain CCSN explosion energies of $E_{\rm SN} \ga 2 \times 10^{51} \erg - 3 \times 10^{51} \erg$ (e.g., \citealt{Gogilashvilietal2021}; section \ref{subsec:AngularMomentumSource}). The most popular extra energy source in these modellings is a rapidly rotating magnetized NS, i.e., a magnetar (e.g., \citealt{Maedaetal2007, Greineretal2015, Metzgeretal2015, Yuetal2017, Marguttietal2018, SuzukiMaeda2021}; for possible SLSN energy sources see, e.g., \citealt{WangWangDai2019RAA}). 

The formation of a magnetar is also the expectation in the jittering jets explosion mechanism when the precollapse core is rapidly rotating. However, in the jittering jets explosion mechanism whenever there is an energetic magnetar there are also energetic jets that are likely to deposit more energy to the ejecta than the magnetar does \citep{Soker2016Magnetar, Soker2017Magnetar, SokerGilkis2017}. 
Following these studies \cite{Shankaretal2021} discuss accretion and jets' launching by a magnetar in broad-lined Type Ic CCSNe. 
 
Indeed, studies that do not include jets encounter problems in a large fraction of the SLSNe they try to fit. \cite{Nicholletal2017b} fit lightcurves of 38 SLSNe with magnetars. \cite{SokerGilkis2017} find that in about half of these fittings the CCSN explosion energy, before any magnetar powering, must be $E_{\rm SN}  > 2 \times 10^{51} \erg$. Because this is a larger energy than what the delayed neutrino mechanism can account for \cite{SokerGilkis2017} conclude that jets exploded these SLSNe. The recent magnetohydrodynamic simulations by  \cite{Reichertetal2022} further support the claim of \cite{SokerGilkis2017} that jets explode SLSNe.  
 
In yet another study \citep{Soker2022LSNe} I examine the modelling of the lightcurves of 40 LSNe by \cite{Gomezetal2022}. For their fitting \cite{Gomezetal2022} consider in addition to the explosion energy itself the contribution by a magnetar and by helium burning. I find that in 10 LSNe that \cite{Gomezetal2022} fit the total energy of these two extra energy sources is larger than the ejecta kinetic energy that they fit. In 8 LSNe the total energy of the delayed neutrino explosion mechanism and these two extra sources combined is smaller than the fitted kinetic energy. Instead, I propose in that paper that jets play an important role in powering the explosion and lightcurve of LSNe, as I claim for SLSNe, and actually for most CCSNe. 

In a recent paper \cite{Kangasetal2022} analyze 14 hydrogen-rich SLSNe and try to fit their lightcurves with either a magnetar or with an ejecta-CSM collision. In half of the SLSNe that they fit by magnetar powering the initial energy of the magnetar is very large (spin period  $P_{\rm mag} \le 0.002 \s$) such that the final mass accretion onto the NS was most likely through an accretion disk that launched more energetic jets even \citep{Soker2017Magnetar}. In half of the cases that they fit with ejecta-CSM collision the kinetic energy and radiation combined carry more energy than what the delayed neutrino explosion mechanism can supply (section \ref{subsec:AngularMomentumSource}), again requiring jets to explode these SLSNe. 
I claim that jets played a dominant role in the explosion of all these hydrogen-rich SLSNe. 

The same arguments hold for the fitting by \cite{Chenetal2022} of the lightcurves of 70 hydrogen-poor SLSNe. They fit by either a magnetar model or an ejecta-CSM interaction + $^{56}$Ni. I find that about $33 \%$ of their magnetar fittings have $P_{\rm mag} \le 0.002 \s$. As I argued above \citep{Soker2017Magnetar} such a rapidly rotating magnetars are likely to be spun-up during the last phase of their formation by an accretion disk/belt that launches jets. I argue that even the slower magnetars in their fitting most likely launched jets at explosion. From the 70 lightcurves that \cite{Chenetal2022} fit 14 do much better with  ejecta-CSM powering of the lightcurve than with magnetars. In 7 out of these 14 SLSNe the kinetic energy of the ejecta is much larger than what the delayed neutrino explosion mechanism can supply, e.g., 
$E_{\rm kin} \simeq 2.5 \times 10^{52} \erg$ and $E_{\rm kin} \simeq 2 \times 10^{52} \erg$ for ZTF18aaisyyp and ZTF18aajqcue, respectively. Clearly if these models are correct jets exploded these SLSNe. Again, I argue that jets exploded all these SLSNe.   

In a recent study \cite{Eigeretal2021} find no correlations between the properties of CCSN SNRs and the magnetar and pulsar they host. I argue that their findings are compatible with my claim that jets supply most of the explosion energy and the kinetic energy of the ejecta rather than the magnetars they host.   

Most CCSNe are not bipolar. \cite{Lazzatietal2012} showed in their two-dimensional hydrodynamical simulations of relativistic jets that only in cases of weak bipolar jets the outcome is dynamically indistinguishable from regular CCSNe. I expect in those cases to have jittering jets rather than fixed-axis jets. 
\cite{Barnesetal2018} conduct two-dimensional relativistic hydrodynamics simulations and argue that relativistic jets can explain broad-lined SNe Ic. \cite{Shankaretal2021} reach similar conclusions with two-dimensional relativistic hydrodynamics simulations.
However, in a recent study \cite{Eisenbergetal2022} find that collimated jets do not produce outflows that are consistent with observations of CCSNe. Instead, they argue, in gamma ray bursts there are narrow jets that produce the gamma emission, and a much wider outflow that explodes the star, including a wide cocoon (e.g., \citealt{Paisetal2022}). \cite{WangYetal2022} continue this idea and discuss jets with large opening angles of about tens of degrees to power the CCSN that accompanied the gamma ray burst GRB 171205A. 
Indeed, the general jet feedback mechanism (in CCSNe, in clusters of galaxies, and in binary stars) operate most efficiently with wide jets or with precessing/jittering jets (for review see \citealt{Soker2016Rev}). I expect that in most bipolar CCSNe the exploding jets are not relativistic. 

For comparison, I comment on the NS activity at birth of most CCSNe, which are not bipolar. \cite{Beniaminietal2019M} estimate that a fraction of $0.4^{+0.6}_{-0.3}$ of NSs are born as magnetars. Most of these have long spin periods, 
$P_{s} \approx 0.01-0.1 \s$, and their energy does not dominate nor the explosion nor the lightcurve. \cite{GofmanSoker2020} estimate the spin period of the NS remnants of  CCSNe with explosion energies of $10^{50}-10^{51} \erg$, which are the majority of CCSNe,  to be in the range of $P_{s} \approx 0.01-0.1 \s$. We expect many of them to have medium to strong magnetic fields as magnetic fields are required for jet-launching. Namely, the NS with strong magnetic fields are born as magnetar, but with low energy because of their relatively slow rotation. In this regard the jittering jets explosion mechanism is compatible with the finding of  \cite{Beniaminietal2019M}.

\textbf{$\bigstar$ Summary of section \ref{subsec:SuperluminousSNe}.}
Jets must play an important role in powering the explosion and lightcurves of LSNe and SLSNe, most likely the dominant role. Modelling of SLSN lightcurves must include jets, even when they consider an energetic magnetar. In most cases the jets have a fixed-axis or a small jittering around a fixed axis. The fixed axis results from a rapid precollapse core rotation. A close binary companion (a main sequence star, an NS, or a BH) might spin-up the core to the required rapid rotation. 

\subsection{The role of the viewing angle} 
\label{subsec:Viewingangle}
In CCSNe with bipolar ejecta, i.e., a morphology of two polar lobes with an equatorial waist between them, the viewing angle influences the observed lightcurve.  

\cite{KaplanSoker2020b} built simple bipolar CCSN ejecta models and examined the basic structure of the lightcurve for an observer in and near the equatorial plane. They built the ejecta from an equatorial ejecta and faster polar jet-inflated ejecta. At early times the polar lobes are optically thick, and because of their faster polar velocity the polar photosphere grows faster than the photosphere near the equatorial plane. The outcome is a CCSN that is more luminous than a similar spherical explosion. The jets supply the extra energy to inflate and accelerate the polar lobes and to the extra radiation. As the polar ejecta expands faster, its optical depth decreases faster and at later times the polar photosphere decreases faster than the equatorial photosphere, and eventually the equatorial ejecta hides the polar photosphere from an observer near the equatorial plane. This leads to a rapid luminosity decline, and even an abrupt decline in the lightcurve. 
 
\begin{figure}[ht!]
\centering
\includegraphics[trim=0.95cm 12cm 3.5cm 0.0cm, scale=0.5]{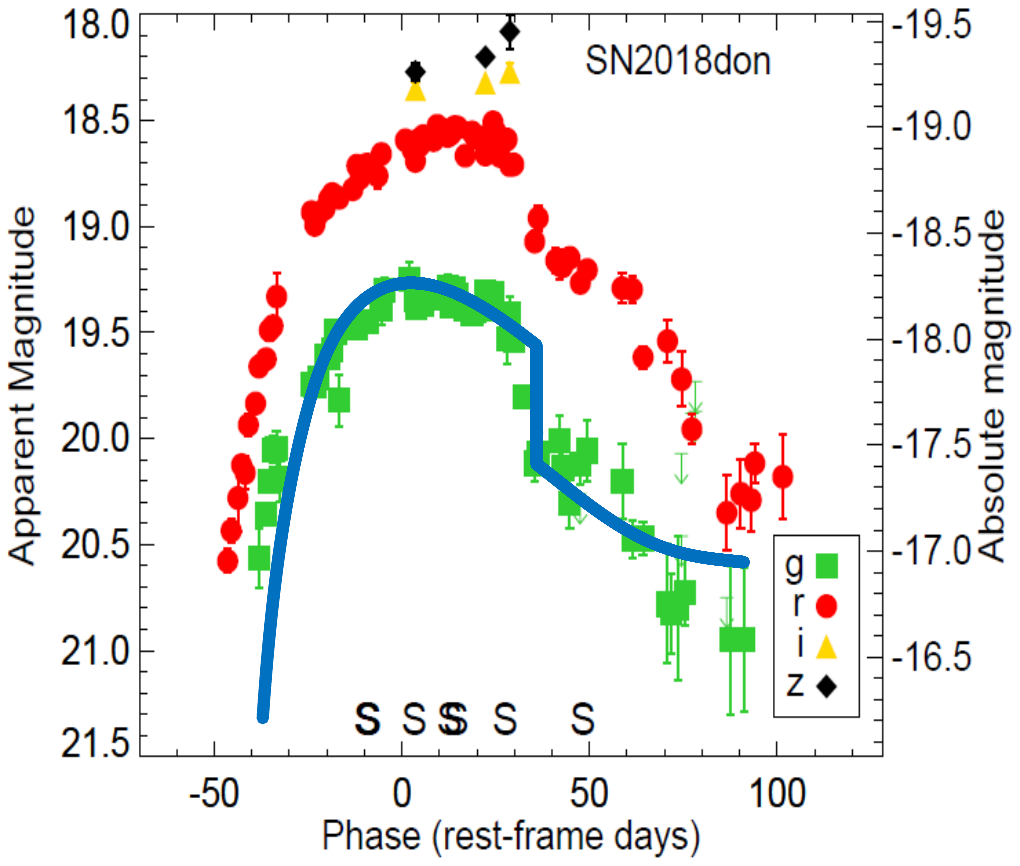}
	\caption{Fitting the lightcurve of a LSNe with bipolar ejecta (figure from \citealt{KaplanSoker2020b}). The green and the red  points are the lightcurves of SN~2018don in the g-band and r-band, respectively, from \cite{Lunnanetal2020}. The blue line is the lightcurve from a toy model by \cite{KaplanSoker2020b} where more details can be found.  
	}
	\label{fig:SN2018don}
\end{figure}

\cite{KaplanSoker2020b} could fit with their toy model the abrupt decline in the light curve of SN~2018don. In Fig. \ref{fig:SN2018don} I present their fitting to the light curve of SN~2018don. In a recent study \cite{AkashiMichaelisSoker2022} show with three dimensional hydrodynamical simulations that bipolar explosions can indeed explain such lightcurves. I suggest \citep{Soker2022LSNe} that bipolar explosion might explain the lightcurve of the stripped-envelope SLSN SN~2020wnt that has a `knee', because the bipolar explosion can account also for that (for observations see \citealt{Tinyanontetal2021, Gutierrezetal2022, Tinyanontetal2022}). I will return to LSNe and SLSNe in section \ref{sec:PostExplosion} where I discuss post-explosion jets. 

Consider fast blue optical transients (FBOTs), and in particular AT2018cow-like FBOTs that are bright transients (e.g., \citealt{Marguttietal2019}) that have only few days rise time to peak emission  (e.g., \citealt{Prenticeeta2018, Perleyetal2019}). They display hydrogen lines (e.g., \citealt{Marguttietal2019}), and show high velocities of $\ga 0.1 c$ and a total kinetic energy of $\simeq 10^{51}-10^{52} \erg$ (e.g., \citealt{Coppejansetal2020}). 

\cite{SokeretalGG2019} and \cite{ Soker2022FBOT} suggest that AT2018cow-like FBOTs are classes of CEJSNe and CEJSN-impostors, respectively (section \ref{sec:CEJSNe}), for which the viewing angle play a crucial role. \cite{Metzger2022FBOT} suggests a version of the CEJSN scenario of \cite{SokeretalGG2019} where the core-NS/BH merger takes place $>100 \yr$ post CEE. In the CEJSN scenario for FBOTs the fast outflow is accounted for by a fast polar outflow and an observer away from the equatorial plane.   
The same holds for the bipolar CCSN explosion model that \cite{Gottliebetal2022cow} propose for AT2018cow-like FBOTs. In their model the jets interact with the stellar envelope. \cite{KashiyamaQuataert2015} propose a model where  the progenitor collapses to form a BH that accretes mass via an accretion disk that power the FBOT by a disk wind. Similarly, \cite{Tsunaetal2021} proposed a CCSN with a bipolar outflow from a BH accretion disk to account for the fast-rising transient AT2018lqh \citep{Ofeketal2021}.
All these models, whether CEJSNe or CCSNe, are bipolar (see also \citealt{Guarinietal2022}), and viewing angle is important. FBOTs are likely observed away from the equatorial plane.

\textbf{$\bigstar$ Summary of section \ref{subsec:Viewingangle}.} Viewing angle influences the lightcurves of bipolar explosions. Bipolar explosions that we observe from the equatorial plane might explain some puzzling features in some CCSN lightcurves. FBOTs are a class of bipolar explosions of CCSNe and/or CEJSNe, that are likely observed along or close to the polar axis.  

\section{Common envelope jets supernovae (CEJSNe)} 
\label{sec:CEJSNe}

\begin{figure*}
\centering
\includegraphics[trim=5.95cm 2.0cm 3.5cm 0.0cm, scale=0.76]{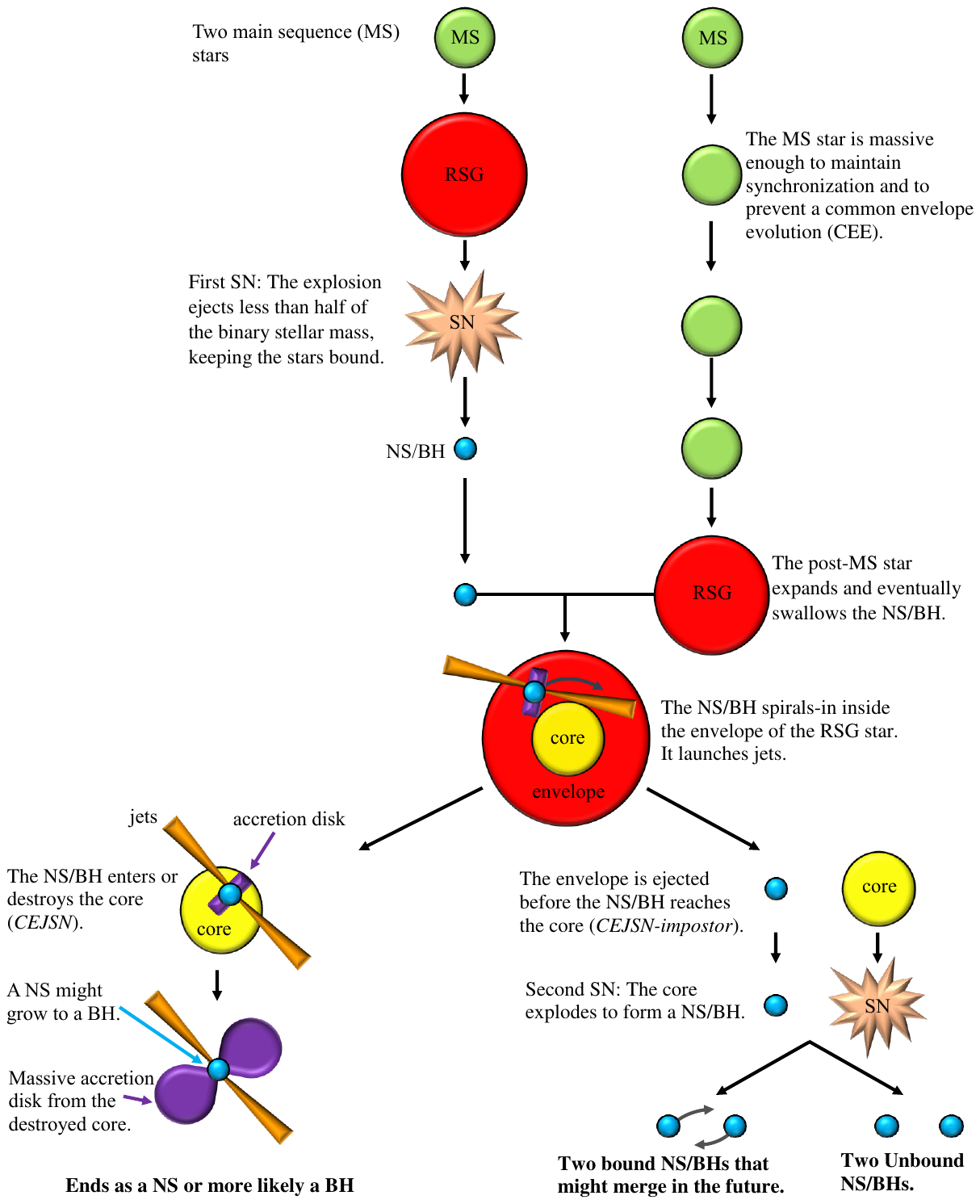}
	\caption{A schematic drawing (not to scale) of the main CEJSN evolutionary phases. Abbreviation. MS: main sequence; NS/BH: a neutron star or a black hole. RSG: red super giant; SN: supernova. 
	}
	\label{fig:CJSNe}
\end{figure*}

A CEJSN is a transient event with typical time scales like CCSNe or longer and with energies like those of CCSNe or larger that is powered by jets that a NS or a BH (NS/BH) launch as they accrete mass from the envelope and then core of a RSG (e.g, \citealt{SokeretalGG2019, GrichenerSoker2019a, Schroderetal2020, GrichenerSoker2021}; for studies of NS/BH merger with the RSG core that do not emphasize jets see, e.g., \citealt{FryerWoosley1998} and \citealt{Chevalier2012}). 
They therefore might mimic (peculiar) CCSNe. 
In cases where the NS/BH does not enter the core, i.e., it enters and then exits the RSG envelope or it removes the entire envelope in a common envelope evolution (CEE) and avoids entering the core, the event is a CEJSN-impostor (e.g., \citealt{Gilkisetal2019, LopezCamaraetal2019, LopezCamaraetal2020MN, Gricheneretal2021}).
However, in this review I will use CEJSN to refer to CEJSN impostors as well as CEJSNe. 
Some CEJSNe involve triple-star interaction (e.g., \citealt{Soker2021Triple, Soker2021NSNS, AkashiSoker2021}). I schematically present the main phases of the CEJSN evolution in Fig. \ref{fig:CJSNe}.

Numerical simulations show that the jets regulate their power because they remove mass from their vicinity and reduce the density, hence the accretion rate and the jets' power (e.g., \citealt{LopezCamaraetal2019, Gricheneretal2021, Schreieretal2021, Hilleletal2022}). This is the jet feedback mechanism in a CEE (e.g., \citealt{Soker2016Rev}). 
Even jets from a main sequence star might efficiently expel mass from the giant envelope (e.g., \citealt{Shiberetal2019}), and more so the energetic jets that a NS/BH launches (\citealt{Schreieretal2021, Hilleletal2022}). 
This efficient mass removal might increases the CEE efficiency parameter to values of $\alpha_{\rm CE} > 1$ as some scenarios demand (e.g. \citealt{Fragosetal2019, Zevinetal2021, Garciaetal2021}).

Two key processes allow energetic CEJSNe. (1) An efficient cooling of the mass that the NS/BH accretes by neutrino emission when the accretion rate is sufficiently high, $\dot M_{\rm acc} \ga 10^{-3} M_\odot \yr^{-1}$ \citep{HouckChevalier1991, Chevalier1993, Chevalier2012}. (2) A density gradient in the RSG envelope and the RSG core that implies that the accreted mass has a specific angular momentum that forms an accretion disk around the NS/BH (e.g.,  \citealt{ArmitageLivio2000, Papishetal2015, SokerGilkis2018}). 

CEJSNe might account for some enigmatic transients that at first are classified as CCSNe, e.g. the unusual gamma-ray burst GRB~101225A for which \cite{Thoneetal2011} suggested the merger of a NS with a helium star. 
\cite{SokerGilkis2018} proposed a CEJSN event to explain the puzzling iPTF14hls transient \citep{Arcavietal2017} and similar transients, e.g.,  SN~2020faa \citep{Yangetal2021}. I will return to iPTF14hls in section \ref{subsubsec:iPTF14hls}. FBOTs (section \ref{subsec:Viewingangle}) and in particular AT2018cow-like FBOTs might be CEJSNe \citep{SokeretalGG2019, Metzger2022FBOT} or CEJSN-impostors \citep{Soker2022FBOT}. 
\cite{Schroderetal2020} proposed that the transients SN1979c and SN1998s were CEJSN events. \cite{Dongetal2021} proposed the CEJSN scenario for the luminous radio transient VT~J121001+495647. 

CEJSN events and CCSNe share the same main  powering mechanism that is the launching of jets by a NS/BH, an old NS/BH companion in CEJSN events and a newly born NS/BH in CCSNe. Therefore, it is not surprising that CEJSNe might mimic CCSNe. Some CEJSNe might be more energetic and/or have a slower declining lightcurves than typical CCSNe. Because of the binary interaction CEJSN events cover a more extended range of explosion properties (e.g., see table  1 of \citealt{SokeretalGG2019}). 
 
Triple-star CEJSNe add to the rich variety of properties, e.g., a tight binary system of a NS and a main sequence star enters the envelope of a RSG and merge inside the RSG envelope \citep{Soker2021Triple}, or a tight binary system of two NS/BHs that enters the RSG envelope and ends in one of several CEJSN channels (table 1 of \citealt{Soker2021NSNS}), like a NS-NS merger inside the envelope \citep{AkashiSoker2021}.

In their population synthesis study  \cite{Schroderetal2020} estimate that the rate CEJSN events where a NS/BH enters the core of an RGB star is $\approx 0.5 \%$ of the rate of CCSNe. About half the cases are with a NS star companion and about half with a BH companion. 
The CEJSN-impostor events, where the NS/BH does not enter the core, amount to $\approx 2 \%$ of the rate of CCSNe. 

\textbf{$\bigstar$ Summary of section \ref{sec:CEJSNe}.} Jets that a NS/BH launches power CEJSNe, much as I argue for CCSNe. The main differences are that in CEJSNe the NS/BH are old and the source of the accreted mass is the envelope and then the core inside which the NS/BH orbits, i.e., a CEE. Because of the similar powering mechanism  CEJSNe mimic CCSNe and in particular bipolar CCSNe, but have some other properties, like they might last longer and contain more energy. 
The rate of CEJSNe is only few per cent of the CCSN rate. Some CEJSN-impostors might end as a close NS/BH-NS/BH binary system that much later suffers merger due to gravitational wave emission. In other words, many of the gravitational wave sources that evolved through a CEE have experienced a CEJSN-impostor event. 

\section{Shaping the CSM} 
\label{sec:PreExplosionCSM}

\subsection{Core collapse supernovae} 
\label{subsec:1987A}

\begin{figure*}
\centering
\includegraphics[trim=3.00cm 0.0cm 3.5cm 0.0cm, scale=0.75]{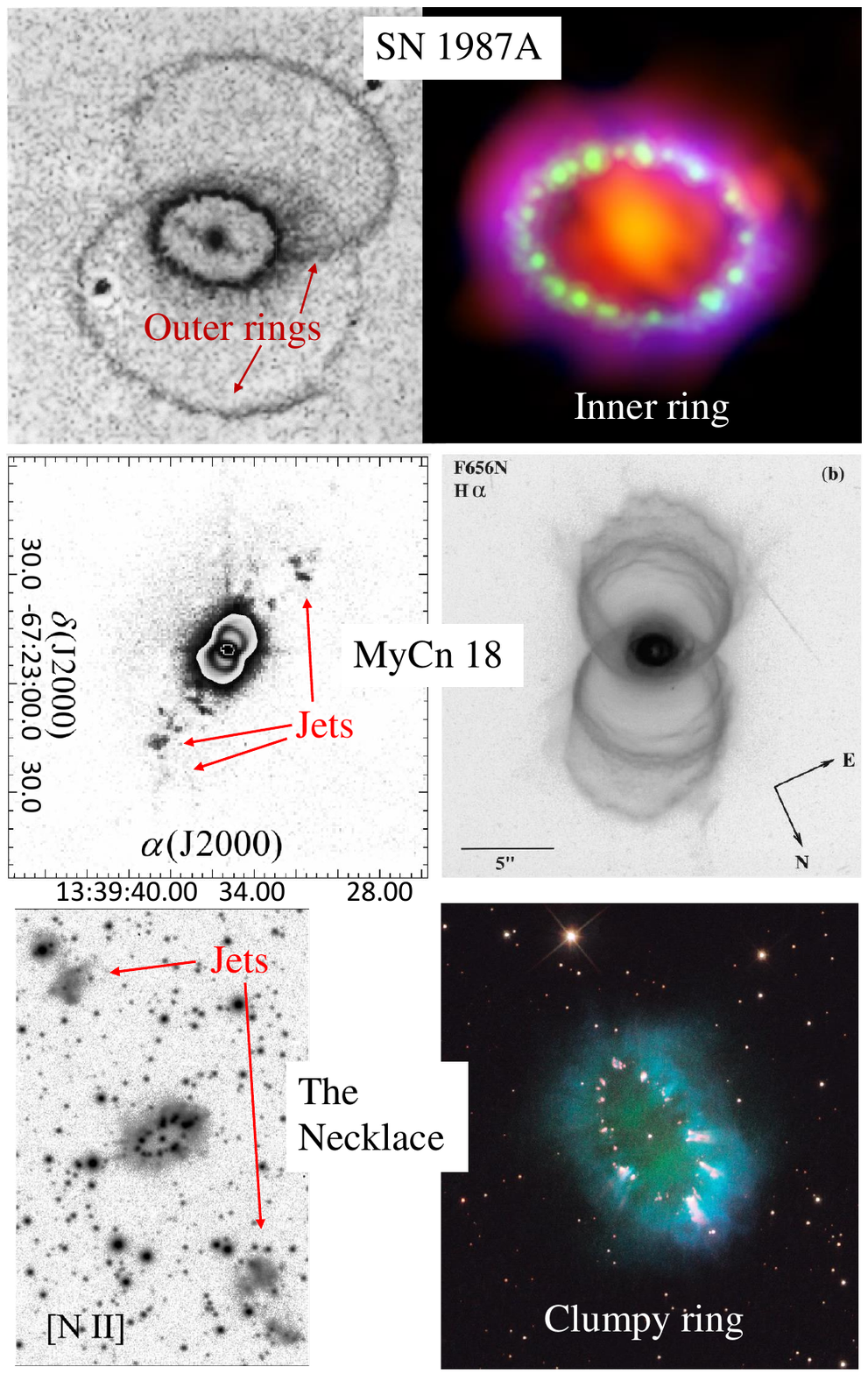}
	\caption{Comparing the morphology of the three CSM rings of SN 1987A to the morphologies of two planetary nebulae. The right column are zoomed-in images of the left column images, but do not have the same orientation. Details of the images that are not of concern to this study can be found in the sources of the images. 
	Credits: SN 1987a: Left: \cite{Burrowsetal1995}; Right: X-ray: NASA/CXC/SAO/PSU/K, \cite{Franketal2016}; Optical: NASA/STScI; Millimeter: ESO/NAOJ/NRAO/ALMA. MyCn~18: Left: \cite{OConnoretal2000}; Right: \cite{Sahaietal1999}. The Necklace nebula (IPHASX J194359.5+170901): Left: \cite{Corradietal2011}. Right:  NASA/ESA/the Hubble Heritage Team (STScI/AURA).
	}
	\label{fig:1987A}
\end{figure*}

I start with the three rings of SN~1987A that I present in Fig. \ref{fig:1987A}. The ejecta of SN 1987A has been strongly colliding with the inner ring for about 23 years (e.g., \citealt{Franssonetal2013, Franketal2016, Larssonetal2019}; see \citealt{McCrayFransson2016} for a review). The rings were ejected $\approx 20,000 \yr$ before explosion (e.g., \citealt{Panagiaetal1996, McCrayFransson2016}). 

Although there are scenarios for the formation of the three rings of SN 1987A that involve no jets (e.g., \citealt{Soker1999Rings87A,TanaksWashimi2002, Sugermanetal2005, MorrisPodsiadlowski2009}), I attribute their shaping to jets \citep{Soker2002Rings}. The supporting argument comes from the morphological similarities to those of some PNe. In Fig. \ref{fig:1987A} I present these similarities with two PNe.
Note that these jets are likely to be wide, like jets from post-asymptotic giant branch binary systems (e.g., \citealt{Bollen2022}; for review emphasizing wide jets see \citealt{Soker2016Rev}).

The PN MyCN~18 and SN 1987A share a bipolar structure of pairs of opposite outer rings, one pair in SN 1987A and several pairs in MyCn~18. The deep image of MyCn~18 reveals broken jets at large distances from the bipolar structure and along its symmetry axis. Most likely several jet-launching episodes shaped the several rings of MyCn~18 as a result of the interaction of the jets with a CSM. 
This supports a scenario where pairs of jets shaped the pair of outer rings of SN~1987A, as \cite{AkashiSoker2016Rings} demonstrate in their three-dimensional hydrodynamical simulations. 
  
The Necklace PN and SN 1987A share the structure of a clumpy equatorial ring. Again, the Necklace PN was shaped by jets as the image shows. This suggests that jets shaped also the inner ring of SN 1987A, as \cite{Akashietal2015} show with three-dimensional hydrodynamical simulations. I do not argue that necessarily only jet compressed the equatorial outflow that formed the ring. Equatorial mass ejection during a CEE phase might added mass to the ring. 

The star that launches the jets in the scenario I discuss here is a main sequence companion that entered a CEE with the RSG (then) progenitor of SN 1987A. Before it entered the envelope the companion accreted mass from the RSG envelope via an accretion disk that launched the jets. As it spiralled-in towards the envelope and inside it, the companion spun-up the progenitor of SN 1987a (e.g., \citealt{ChevalierSoker1989}) and diverted its evolution towards a blue giant (e.g., \citealt{Podsiadlowskietal1990}). The companion did not survive the CEE. 

There are other CSM morphologies into which CCSN explosions might occur, e.g., a bipolar CSM around a WR star (e.g., \citealt{Meyer2021}). Therefore, in principle a bipolar CCSNR might result from either jets during the explosion or a bipolar CSM, or both.  

\textbf{$\bigstar$ Summary of section \ref{subsec:1987A}.}
I suggest that the three CSM rings of SN 1987A were ejected by a binary system that entered a CEE $\approx 20,000 \yr$ before explosion, and that a main sequence  companion to the progenitor of SN 1987A launched jets that shaped the rings. The interaction of the jets with the CSM gave rise to a transient event, an intermediate luminosity optical transient (ILOT). 

\subsection{Shaping ears in type Ia supernovae} 
\label{subsec:TypeIaSNe}

This section is the only one in this review where I refer to type Ia supernovae (SNe Ia). Jets do not power SNe Ia and so I refer only to the shaping. As well, I will not get into the question of which SN Ia scenarios allow ears (for that see table 1 in \citealt{Soker2019Rev}). 

\begin{figure*}
\centering
\includegraphics[trim=3.00cm 1.6cm 3.5cm 0.0cm, scale=0.72]{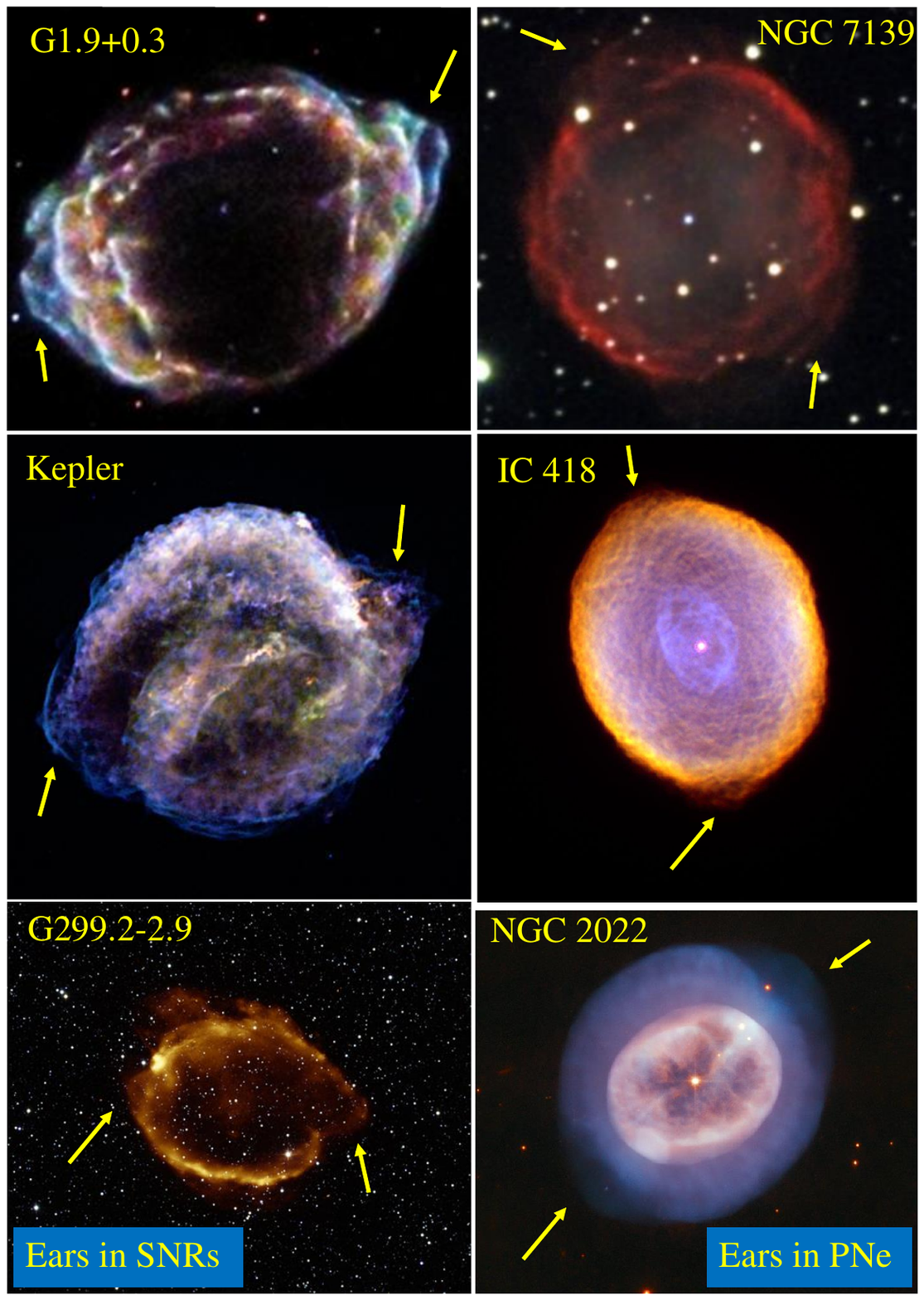}
	\caption{Comparing `ears' in three SNRs Ia with ears of three PNe. Each yellow arrow points at an ear. See NGC 40 and NGC 3918 in fig. \ref{Fig:JetsPNeArcs} for PNe with larger ears and NGC 3132 in Fig \ref{Fig:JetsPNeEars} for asymmetrical ears (I marked the ears there `jet opening' to indicate that we identify the jets in some of these PNe). I adopt the view that these and similar SNRs Ia are \textit{SNIPs} (for SNe inside PNe). 
	Credits: G1.9+0.3: \cite{Borkowskietal2014}; Kepler: \cite{Reynoldsetal2007}; G299.2-2.9: \cite{Parketal2007} (for SNR images see also the Chandra homepage); IC 418: NASA/ESA and The Hubble Heritage Team (STScI/AURA); NGC 7139: KPNO/NOIRLab/NSF/AURA/Gert Gottschalk and Sibylle Froehlich/Adam Block; NGC 2022: ESA/Hubble and NASA, R. Wade.
	}
	\label{fig:TypeIaSNe}
\end{figure*}

Many SNRs of SNe Ia (SNRs Ia) have the structure of two opposite ears. In Fig. \ref{fig:TypeIaSNe} I present three SNRs Ia with ears alongside three PNe with small ears; some PNe with large ears are in Figs. \ref{Fig:JetsPNeEars} and \ref{Fig:JetsPNeArcs}. \cite{TsebrenkoSoker2015a} and \cite{Chiotellisetal2021} give more examples of SNRs Ia with ears. 
 
I will follow \cite{TsebrenkoSoker2015a} and take the view that jets shape ears in SNRs Ia. \cite{TsebrenkoSoker2013} simulated the shaping of ears by either a spherical SN Ia exploding into a CSM with ears, or by jets in the SN Ia explosion itself that expand into a spherical CSM. The SN Ia explosion simulations by \cite{Peretsetal2019}, where a CO white dwarf tidally destroys a HeCO white dwarf, yield faster bipolar outflow that might shape ears. Their study show that some SN Ia explosion scenarios might shape jets. However, I will take the view that SNe Ia are more or less spherical (beside maybe some clumps) and that the ears result from the CSM that the progenitor of the SN Ia blew hundreds of thousands or years to thousands or years before explosion. The CSM is actually a proto-PN \citep{Cikotaetal2017}, a PN, or a remnant of a PN, and these are SNe Ia inside PNe, which are termed \textit{SNIPs} \citep{TsebrenkoSoker2015a}. I crudely estimated that about half of SNe Ia are SNIPs \citep{Soker2022CEEDDT}.

It is important to emphasize that here I argue that the jets where launched during the formation phase of the PN that preceded the SN Ia explosion. The SN Ia itself has a large-scale spherical structure. It explodes into a CSM with ears, a prot-PN, a PN, or an old PN.  

In the model I adopt here the ears are along the symmetry axis of the SNR.  Some models for ear formation take them to be projections of an equatorial ring/torus. \cite{Chiotellisetal2021} simulate the formation of ears in SNIPs, but they take the ears to be in the equatorial plane and do not include jets (see also \citealt{Chiotellisetal2020Galax}). \cite{Burkeyetal2013} also take the ears of the Kepler SNR  to be in the equatorial plane. I note that \cite{SunChen2019} find that the ears in the Kepler SNR consist mainly of Si-rich and S-rich ejecta, thus, they argue, favoring shaping by jets that are related to the explosion, rather than ears from a preceding planetary nebula. This requires further study.  

In the Kepler SNR (e.g., \citealt{Kasugaetal2021}) and in the SNR G1.9+0.3 (e.g., \citealt{Borkowskietal2017}) the ejecta interacts with a CSM. \cite{TsebrenkoSoker2015G19} simulates the formation of the complicated ears SNR G1.9+0.3 by adding clumps to the ears in the PN. In the SNR G299.2-2.9 the ears are less symmetric, with a clear elongation only in one side.
\cite{Postetal2014} consider the possibility that G299.2-2.9 interacts with a CSM, but also argues that it is unlikely to be a SNIP because of only one-sided ear. 
However, I note that some PNe do have very unequal ears. A good example is the Webb’s MIRI image of NGC 3132 that I present in the right-most panel in the second row of Fig. \ref{Fig:JetsPNeEars} (for an analysis of this image see \citealt{DeMarcoetal2022}) .

\textbf{$\bigstar$ Summary of section \ref{subsec:TypeIaSNe}.} Some SNRs Ia have ears that I take to indicate shaping by jets. Although in principle the jets might be part of the explosion process, I consider more likely the explosion of a large-scale spherical SN Ia inside a CSM that is a PN (SNIP). The ears result from the shaping of the PN by jets. 

\section{Jet-driven pre-explosion outbursts} 
\label{sec:PreExplosionOutbursts}

This section deals with pre-explosion jets in binary systems where a companion to the progenitor of the CCSN (or the CEJSN) launches the jets as it accretes mass from the progenitor. 

Observations show that some CCSNe and some other transients experience outbursts years to days before explosion. These are termed precursors (e.g., \citealt{Tsunaetal2022} for a recent lightcurve calculations). Prominent is the class of SN~2009ip-like (2009ip-like; e.g., \citealt{Smithetal2010, Smithetal2014, Fraseretal2013, Mauerhanetal2013, Pastorelloetal2013, Grahametal2014, Smithetal2022}) transients that have precursors, fast rising and declining lightcurves and bumps during the decline phase. 
Some other members of this group are 
SN~2010mc (e.g., \citealt{Ofeketal2013, Smithetal2014}), LSQ13zm \citep{Tartagliaetal2016}, SN~2013gc \citep{Reguittietal2019},  SN~2015bh (e.g., \citealt{EliasRosaetal2016, Thoneetal2017}), AT~2016jbu (e.g., \citealt{Boseetal2017ATel, Kilpatricketal2018, Brennanetal2022a}), SN~2016bdu (e.g., \citealt{Pastorelloetal2018}), and SN 2019zrk \citep{Franssonetal2022, Strotjohanntal2021}. 
In their study of SN 2019zrk \cite{Franssonetal2022} point out that 2009ip-like transients might emit a substantial fraction of the radiation in the UV and in X-rays as in SN~2009ip \citep{Marguttietal2014}.
All these properties are not easy to explain by CCSNe, unless some other ingredients are added. 
 
It is not clear what fraction of 2009ip-like transients are true CCSNe (e.g., \citealt{Smithetal2014}) and what fraction are CEJSNe (e.g., \citealt{Gilkisetal2019, Schroderetal2020}). My view is that this group of transients bridges in its properties and powering CCSNe and CEJSNe. In other words, some of them might be true CCSNe and some might be CEJSNe, but in all of them jets play major roles before, during, and after the explosion, and the ejecta has a highly non-spherical morphology, most likely bipolar. 

The quest for an explanation to the pre-explosion outbursts encounters some challenges. 
\cite{QuataertShiode2012} and \cite{ShiodeQuataert2014} suggested that waves that the vigorous convection motion in the core, from the phase of core carbon burning until core collapse, can excite waves that propagate to the envelope.
\cite{SokerGilkisDynamo} proposed that the vigorous convection can set a magnetic activity where magnetic flux tubes that buoy to the envelope deposit energy to the envelope. 
The dissipation of the wave energy and/or magnetic energy in the envelope heats the envelope and inflates it (e.g., \citealt{Fuller2017, FullerRo2018, WuFuller2021}).

However, models that are based on only wave energy (or magnetic energy) encounter two problems. The first one is that waves cause mainly envelope inflation rather than mass ejection (e.g., \citealt{McleySoker2014, OuchiMaeda2019, WuFuller2022}), and cannot explain dense CSM, e.g., as \cite{Franssonetal2022} mention for SN 2019zrk. The three-dimensional simulations by \cite{Tsangetal2022} yield somewhat more ejected mass, but cannot solve the problem. The second challenge, as mentioned for example by \cite{Franssonetal2022}, is that there are observations of fast, velocities of $\simeq 10^4 \km \s^{-1}$, outflows in pre-explosion outbursts, like in SN 2009ip (e.g., \citealt{Pastorelloetal2013}).

The similarity of some pre-explosion outbursts to the outbursts of ILOTs, e.g., as \cite{Smithetal2010} and \cite{SokerKashi2013} discussed for SN~2009ip and \cite{Brennanetal2022b} discussed for AT~2016jbu, suggests that jets power these outbursts much as they power ILOTs (other names include gap objects, luminous red novae and intermediate luminosity red transients). Namely, the main energy source of mass ejection during pre-explosion outbursts is accretion onto a compact companion that launches jets. The companion can be a main sequence star, a WR star, or a NS/BH.
Some ILOTs must be powered by jets because of their bipolar morphology, e.g.,  Nova 1670 (CK Vulpeculae; \citealt{Kaminskietal2021}), and I take the view that most ILOTs are powered by jets (e.g., \citealt{Soker2020ILOTs}). 

Indeed, jets can solve the two problems that models that are based on wave energy only encounter (for a more detailed discussion see \citealt{Soker2022zrk}). \cite{McleySoker2014} already found that the energy that waves are likely to deposit to the RSG envelope before explosion mainly causes envelope expansion, but not much mass ejection. Their conclusion was that a mass-accreting companion powers the outbursts by launching jets. The waves inflate the envelope that in turn transfers mass to the more compact companion via an accretion disk that launches energetic jets.  \cite{DanieliSoker2019} studied this process in cases where the companion is a NS. 

Jets can also account for high velocities of a fraction of the pre-explosion ejecta. 
Hydrodynamical simulations showed that even jets from main sequence or WR companions that expand at velocities of $v_{\rm jet} \approx 2000-3000 \km \s^{-1}$ can accelerate gas to velocities of $\ga 10,000 \km \s^{-1}$ when they interact with the CSM. 
\cite{TsebrenkoSoker2013JetsIP} showed this process with  parameters that fit the expected pre-explosion of SN~2009ip and \cite{AkashiKashi2020} obtained such high velocities when they applied parameters that fit the Great Eruption of Eta Carinae, which is an energetic ILOT event. 
The observations of high pre-explosion velocities is not a problem at all for cases with a NS/BH companion because NS/BHs launch very fast and energetic jets (e.g., \citealt{Gilkisetal2019}). 

\textbf{$\bigstar$ Summary of section \ref{sec:PreExplosionOutbursts}.} 
Although the energy that core-excited waves (and possible magnetic flux tubes that buoy out) carry to the envelope might cause some CCSN progenitors to expand and lose some mass before explosion, energetic pre-explosion outbursts (precursors) require jets for their powering. A more compact companion that accretes mass from the inflated envelope of the CCSN progenitor launches these jets.

\section{Post-explosion jet-powering} 
\label{sec:PostExplosion}

\subsection{Post-explosion jets} 
\label{subsec:LateJets}
 
A NS/BH launches post-explosion jets if it accretes mass through an accretion disk after explosion (I do not consider jets from a pulsar). Several processes can lead to post-explosion accretion in CCSNe. 

Many studies refer to post-explosion accretion of fallback gas (e.g, \citealt{Chevalier1995PhR, DellaValleetal2006, Moriyaetal2010, AkashiSoker2022, Pellegrinoetal2022}), including in bipolar jet-driven CCSNe (e.g., \citealt{Tominaga2009}), and in relation to gamma ray bursts (e.g., \citealt{LinJetal2020}). \cite{DexterKasen2013} conduct a thorough study of fallback accretion in CCSNe and discuss up to months-long post-explosion jets that power SLSNe. The processes that they discuss are relevant to this review, and I will not review these processes that cause late fallback mass accretion. I also note that \cite{Metzgeretal2015} study the detail powering by a magnetar and briefly mention that a BH accreting fallback gas has the same effects on the lightcurve. As I already mentioned in this review and in earlier papers, my view is that energetic magnetars must be accompanied by jets that most likely supply more energy to the explosion. Therefore, jets and magnetar powering are not two alternatives. Rather both jets and magnetar operate or jets alone, but not an energetic magnetar alone. 
 
There are other processes that might lead to post-explosion jets in CCSNe, including the accretion of CCSN ejecta gas by a pre-existing NS/BH  (e.g., \citealt{Fryeretal2014, Becerraetal2015, Becerraetal2019, AkashiSoker2020}) and the feeding of the newly born NS/BH by a close main sequence star that the explosion causes its envelope to inflate (e.g., \citealt{Ogataetal2021, Hoberetal2022}), or even a direct collision of the NS with the main sequence star \citep{HiraiPodsiadlowski2022}. In a recent study \cite{Becerraetal2022} find in their simulations of a CCSN with a close NS that both NSs, the old and the newly-born NS, accrete mass within several minutes after the explosion. They, however, do not simulate jets. I expect both NSs to launch jets (e.g., \citealt{AkashiSoker2020}).  

In CEJSNe the post-explosion accretion is more likely even due to the large angular momentum in the system. This leads to fixed-axis jets that might leave equatorial gas bound after the explosion, much as in CCSNe that have rapid pre-explosion rotation and collapse to form a BH (section \ref{subsec:SuperluminousSNe}). 

\textbf{$\bigstar$ Summary of section \ref{subsec:LateJets}.} 
There are several ways to feed a NS/BH through an accretion disk after an explosion. It might be a newly born NS/NH in CCSNe or an old NS/BH in CEJSNe and in rare CCSNe. This feeding requires a large value of specific angular momentum in the system, most likely by a binary system. Either one star survives or both stars do. The same binary system can trigger also pre-explosion jets. Therefore, many explosions that have late jet-powering might also have precursors.    

\subsection{Jet-powered bumps} 
\label{subsec:Bumps}
In this section I argue that some (but not all) bumps in the lightcurves of CCSNe and CEJSNe are powered by jets. As such, studies cannot ignore the role of jets when building models for bumps and the powering by jets should be compared with the powering by other energy sources. This is true in particular for what I term sharp-bumps, which are defined as relatively luminous and short bumps.
 
Many studies refer to magnetar (e.g., \citealt{ChugaiUtrobin2022}) and/or ejecta-CSM interaction powering of bumps (e.g., \citealt{Hosseinzadehetal2022} for a recent study). \cite{Matzgeretal2018} do mention fallback accretion by a magnetar that can change the smooth magnetar power. 

\subsubsection{The toy model} 
\label{subsubsec:ToyModel}

 The collision of a CCSN ejecta with a close CSM channels kinetic energy to thermal energy and then radiation, and as a result of that can delay the lightcurve decline and/or power bumps in the lightcurve (e.g., \citealt{Lietal2020, Fioreetal2021, Sollermanetal2021, Chenetal2022, Zenatietal2022} for some examples from the last three years).
This holds also for 2009ip-like transients (e.g., \citealt{Brennanetal2022a, Franssonetal2022}).
\cite{Gangopadhyayetal2020} discuss ejecta-CSM interaction in the peculiar Type IIn SN 2012ab and mention that the explosion has a jet-structured outflow. This CCSN seems to have clear indications for a jet-driven explosion. 
I will consider the possibility that some (but definitely not all) bumps in the lightcurves of CCSNe and CEJSNe are powered by late jets. 
Alternatively to ejecta-CSM interaction \cite{Moriyaetal2022} suggest that magnetar can power bumps. My view (section \ref{subsec:SuperluminousSNe}) is that any energetic magnetar is accompanied by even more energetic jets. 
 
We do not have yet a model to derive the properties of the jets for a given bump because of the large volume of the parameter space and because the jet-ejecta interaction, which does not have a spherical symmetry, is very complicated. Parameters include the duration of the jet-activity phase, the velocity of the jets, their mass outflow rate, their opening angle, and the morphology of the CSM, which I expect to be non-spherical. Nonetheless, using simple toy models we have derived some plausible set of parameters in two specific cases (sections \ref{subsubsec:iPTF14hls} and \ref{subsubsec:SN2019zrk}). 

\begin{figure}
\centering
\includegraphics[trim=31cm 12cm 31cm 5cm ,clip, scale=0.13]{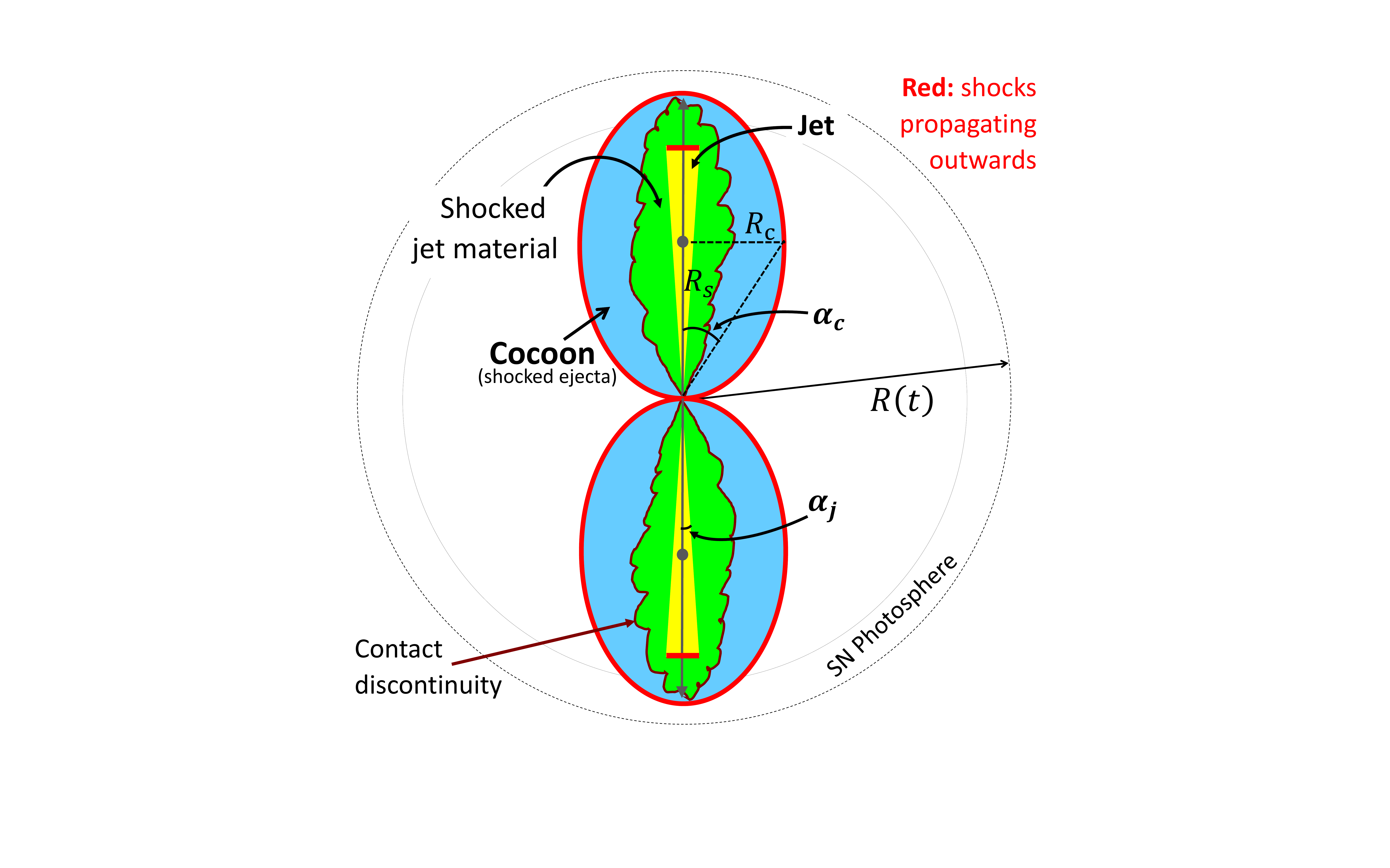}
	\caption{A Schematic illustration (not to scale) of the interaction of the two opposite jets with the SN ejecta--the `mini-explosions' (from \citealt{KaplanSoker2020a}). 
	}
	\label{fig:ToyNoaKaplan}
\end{figure}

\cite{KaplanSoker2020a} built a toy model that gives the timescale and extra luminosity of a jet-driven bump in the lightcurve. The toy model cannot give the details of the extra light curve. The model takes the jet-ejecta interaction to be short and treats it like a `mini-explosion' inside the ejecta, as Fig. \ref{fig:ToyNoaKaplan} schematically shows. There are two opposite mini-explosions for the two opposite jets. The interaction of a jet with the ejecta shocks the jet material and the ejecta. The shocked regions together make the `cocoon'. 
The input parameters of the toy model are the
explosion energy and mass of the ejecta, the energy of the jets, the initial mass in the cocoons, the half opening angle of the jet, the opacity, the location of the jet-ejecta interaction relative to the photosphere, and the time of the jet-ejecta interaction. 

\subsubsection{The sharp bump in the lightcurve of iPTF14hls} 
\label{subsubsec:iPTF14hls}

There are several late peaks in the lightcurve of iPTF14hls \citep{Arcavietal2017, Sollermanetal2019}, which is an  enigmatic transient for its high luminosity, fast outflow, slow decay, and bumps in the lightcurve.
Theoretical scenarios to explain iPTF14hls include powering by a magnetar (e.g., \citealt{Arcavietal2017, Dessart2018, Woosley2018}), a pair instability supernova (e.g., \citealt{Woosley2018, VignaGomezetal2019, WangLJetal2022}), an interaction of the ejecta with a CSM (e.g., \citealt{AndrewsSmith2018, MilisavljeviMargutti2018}), fallback accretion (e.g., \citealt{Arcavietal2017, Wangetal2018, Liu2019}), a CEJSN event (e.g., \citealt{SokerGilkis2018, YalinewichMatzner2019}), models based on late accretion onto a BH (e.g.,  \citealt{Chugai2018}) or a NS (e.g.,e.g., \citealt{Liu2019}), a jittering jets explosion model that forms a black hole \citep{Quataertetal2019}, and wind-driven models \citep{Moriyaetal2020, UnoMaeda2020}. In some of these scenarios jets play the major powering role  (e.g., \citealt{Chugai2018, SokerGilkis2018, Liu2019, Quataertetal2019}).

\cite{KaplanSoker2020a} focused on one bump in the lightcurve of iPTF14hls, the third peak according to the definition of \cite{Wangetal2018}. \cite{Wangetal2018} attributed the powering of the bumps to intermittent accretion of fallback material with a total mass of $\approx 0.2 M_\odot$. However, they could not fit the third peak because it is bright and short, i.e., a sharp bump. They suggested that the third peak is due to a magnetic activity of the NS. 

The third peak has a duration of 30 days and a total extra radiated energy of $E_{\rm rad,b} \simeq 10^{49} \erg$.
\cite{KaplanSoker2020a} found that they could fit the properties of the third peak with their toy model for jets that together have a kinetic energy of $E_{\rm jets} = 0.016  E_{\rm SN}$ and for an initial mass in the two cocoons of $M_{\rm cocoons}=0.027 M_{\rm SN}$, where $E_{\rm SN}$ and $M_{\rm SN}$ are the explosion energy and the ejecta mass, respectively. The last number implies that the initial half-opening angle of the cocoon (see Fig. \ref{fig:ToyNoaKaplan}) is $\alpha_c= 13.2^\circ$ .   
The jet-ejecta interaction time is $t_{\rm j,0} =309~{\rm d}$. About third of the energy of the jets ends in radiation for these parameters, $E_{\rm rad,b} \simeq 0.3 E_{\rm jets}$. This is a high efficiency relative to an ejecta-CSM interaction (section \ref{subsubsec:SN2019zrk}). \cite{KaplanSoker2020a} took the `mini-explosion' location (the place where the jet deposit the energy in the ejecta) to be at half the photosphere radius and the half opening angle of the jets to be $\alpha_j=5^\circ$.

To explain the radiated energy \cite{KaplanSoker2020a} found that the energy in the two jets combined is $E_{\rm jets} =  3.5 \times 10^{49} \erg$. If the jets carry $\simeq10 \%$ of the accretion energy and have an initial velocity of $v_j \simeq 10^5 \km \s^{-1}$, they found that the central object should accrete a mass of  $M_{\rm acc,3} \simeq 0.0035M_\odot$ in this jet-launching episode. 

In principle jets can account for the other peaks in the light curve of iPTF14hls, but for that we should use longer-lasting jets with varying mass loss rates, rather than a short jet-launching episode. 

I do not claim that the above set of jets' parameters is unique, but rather that it is possible to explain some bumps with jets much better than with ejecta-CSM interaction. Definitely there is also the ejecta-CSM interaction that powers the lightcurve in iPTF14hls and many other CCSNe and CEJSNe, but jets can better explain sharp bumps, as I also demonstrate next.  

\subsubsection{The sharp bump in the lightcurve of SN 2019zrk} 
\label{subsubsec:SN2019zrk}

In a recent study \citep{Soker2022zrk} I used the toy model of \cite{KaplanSoker2020a}  (section \ref{subsubsec:ToyModel}) to explain the sharp bump in the light curve of SN 2019zrk. This peak starts at $t \simeq 95~{\rm days}$ (measured from the rise of the main peak), has its maximum luminosity at $t_{\rm bump} \simeq 110~{\rm days}$, declines on a similar time to its rise time, and has a total extra radiation of $E_{\rm rad,b}\simeq 1.8 \times 10^{48} \erg$ \citep{Franssonetal2022}.  

\cite{Franssonetal2022} argued that this bump is powered by ejecta-CSM interaction. However, I argued \citep{Soker2022zrk} that this explanation encounters severe problems. I noted that the spectra during the bump ($t=109~{\rm days}$) is very similar to that before the bump ($t=83~{\rm days}$). This implies that the powering mechanism of the bump should affect both the ejecta photosphere and the H$\alpha$ emitting gas. 
However, the ejecta-CSM interaction takes place at a much larger radius than the photosphere radius at these late times. Therefore, only a small fraction of the energy that the ejecta-CSM collision process emits will reach the photosphere, $f_{\rm ph} \simeq 0.006$. It is unlikely that the ejecta-CSM interaction has sufficient energy to explain this bump. I would expect the ejecta-CSM collision to change the spectrum as it should not have the same spectrum as the photosphere. These arguments of efficiency and spectrum should be examined in any claim that an ejecta-CSM collision powers bumps. 

In applying the toy model \citep{Soker2022zrk} I took the bump's timescale to be its rise time, $t_{\rm toy,b} = 15  {~\rm d}$, and for its luminosity $L_{\rm toy,b} = E_{\rm rad, b}/t_{\rm toy,b}$.
I took the `mini-explosion' to take place at a radius of 0.1 times the ejecta outer radius because the photosphere has already moved deep into the ejecta. For the same reason I took the fraction of the ejecta mass that the two jets interact with to be only $0.02$, $M_{\rm cocoons}=0.02 M_{\rm SN}$. The jets-ejecta interaction time is $t_{\rm j,0}=100~{\rm days}$.
For the ejecta mass I took $M_{\rm SN}=10 M_\odot$, and assumed that the front of the ejecta moves at a faster velocity than the observed velocity after ejecta-CSM collision of $v_{\rm ex} \simeq 1.6 \times 10^4 \km \s^{-1}$ \citep{Franssonetal2022}, such that the explosion energy itself is $ E_{\rm SN} = 0.3 M_{\rm SN} (2 \times 10^4 \km  \s^{-1} )^2 = 2.4 \times 10^{52} \erg$. 
I took the other parameters as in  \cite{KaplanSoker2020a}. 
 
I found that I can account for the timescale and luminosity of the sharp bump of SN 2019ark if the two jets carry a total energy of $E_{\rm 2j} \simeq 0.0012 E_{\rm SN} = 3 \times 10^{49} \erg$. About $6 \%$ of the this energy ends in the extra radiation of the bump. 

Again, this set of the toy model parameters is not unique, but demonstrates the possibility that jets power some bumps. 

\textbf{$\bigstar$ Summary of section \ref{subsec:Bumps}.}
The main conclusion from this entire section is that jets can account for many sharp bumps (but not all) in the lightcurves of CCSNe and other transients, for some of which other powering models encounter problems. Any modelling of bumps should compare jet-powering with ejecta-CSM interaction and magnetar powering. 
   
\subsection{Long-duration post-explosion jets}
\label{subsec:LongLastingjets}

\begin{figure}
 \centering
\includegraphics[trim= 2.6cm 14.8cm 0.0cm 1.5cm,scale=0.53]{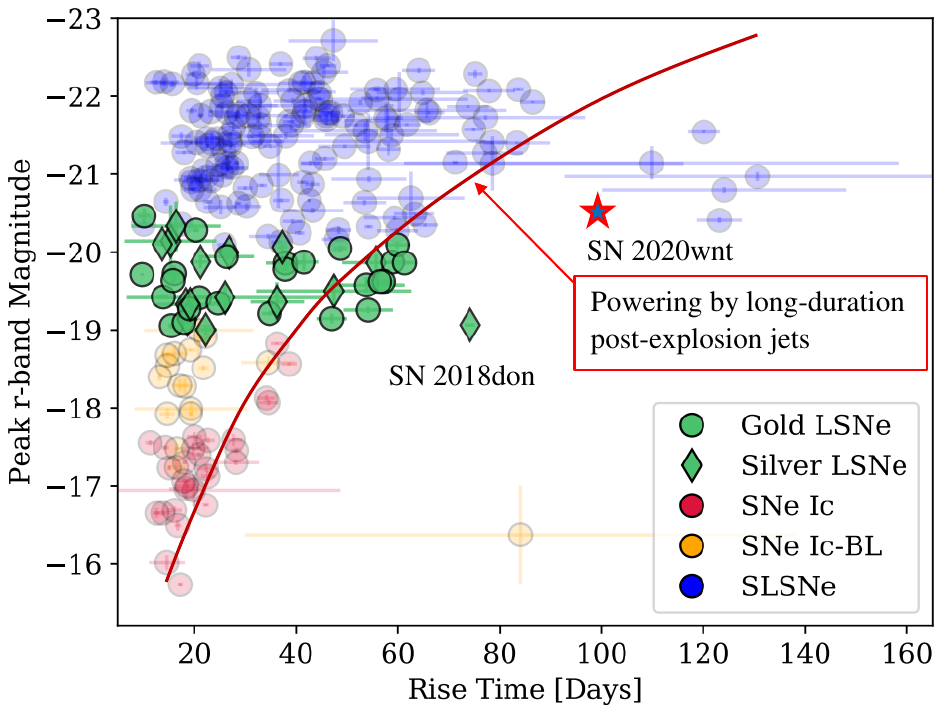}
\caption{A figure from \cite{Gomezetal2022} of CCSNe-I in the plane of peak r-band absolute magnitude versus luminosity rise time with additions of SN~2020wnt and the red line from \cite{Soker2022LSNe}. The inset lists the different CCSN classes.  }
 \label{Fig:LSNeSLSNe}
 \end{figure}

As I commented in section \ref{subsec:LateJets} jets can power the light curve of CCSNe in combination or not with ejecta-CSM interaction and/or an energetic magnetar. In sections \ref{subsubsec:iPTF14hls} and \ref{subsubsec:SN2019zrk} I demonstrated the formation of bumps by short-duration (impulsive) post-explosion jet-launching episodes. I here consider long-duration jets, i.e., jets that the NS/BH launches starting at or immediately after explosion to days to months after explosion. These jets can power LSNe and SLSNe. For that I open by presenting in Fig. \ref{Fig:LSNeSLSNe} a figure from \cite{Gomezetal2022}. In \cite{Soker2022LSNe} I added to this figure SN 2020wnt and a red line that I will explain below. 

There are six SLSNe to the far right of the distribution on Fig. \ref{Fig:LSNeSLSNe}, and one LSN to the far right, SN 2018don. In section \ref{subsec:Viewingangle} I discussed the possibility that strong jets exploded SN 2018don and shaped its ejecta to a bipolar morphology \citep{KaplanSoker2020b}, and that we observe this LSN from near its equatorial plane. \cite{Lunnanetal2020} consider SN~2018don to be a SLSN and not a LSN because extinction implies that its true luminosity is much higher. If this is the case SN~2018don moves higher in Fig. \ref{Fig:LSNeSLSNe} and it becomes a SLSN that is not off to the right. I also mentioned in section \ref{subsec:Viewingangle} that SN 2020wnt might also be a bipolar SLSN that we observed from near its equatorial plane \citep{Soker2022LSNe}. 
 
I now discuss the possibility that long-duration jets power the LSNe and SLSNe with the longest rise time, as I derived in a recent study (\citealt{Soker2022LSNe}, where I give more details). 
In that study I used a toy model that is a modification of the toy model of \cite{KaplanSoker2020a} that I mentioned in section  \ref{subsubsec:ToyModel}. To derive the simplest relation between the rise time and luminosity in this toy model I made several simple modifications and adopted simple assumptions as follows. (1) I assumed that the longest rise time of LSNe an SLSNe for a given luminosity (peak r-band magnitude) is determined by a long jet-activity phase. The interaction process starts early and ends at $t_{\rm j,f}$. (2) I kept the explosion energy and ejecta mass as in the scaling of \cite{KaplanSoker2020a}, i.e., $E_{\rm SN}=2 \times 10^{52} \erg$ and $M_{\rm SN}=10 M_\odot$, respectively. (3) I changed the opacity to $\kappa_{\rm c} =0.1 \cm^2 \g^{-1}$, as a typical value that \cite{Gomezetal2022} list. (4) I assumed that the jets are active for a long time so that the energy the two jets deposit increases with time according to $E_{\rm 2j}(t)=( {t_{\rm j,f}}/{100 {~\rm d}}) {\epsilon_{E,0}} E_{\rm SN}$. 
(5) I assumed that the location of the mini-explosion, namely the radius where the jets deposit their energy inside the ejecta, increases linearly with time and is given by $R_{\rm ME}= ({t_{\rm j,f}}/{100 {~\rm d}}) \beta_0 R_{\rm ej}$, where $R_{\rm ej}$ is the radius of the outer ejecta. 
(6) I kept as in the scaled equation of \cite{KaplanSoker2020a} the values of the mass in one cocoon relative to the ejecta mass $\epsilon_V \equiv M_{\rm 1,cocoon}/M_{\rm SN}$ and the half opening angle of each jet $\alpha_j=5^\circ$.

Using the above modifications and assumptions in the toy model of \cite{KaplanSoker2020a} I obtained \citep{Soker2022LSNe} a relation between the luminosity and rise time of the longest LSNe and SLSNe
\small
\begin{eqnarray}
\begin{aligned} 
& L_{\rm c} \simeq 1.6\times 10^{44} 
\left(\frac{\epsilon_V}{0.067}\right)^{-1}
\left(\frac{\epsilon_{E,0}}{1}\right)
\left(\frac{\sin{\alpha_{\rm j}}}{0.087}\right) \\ & \times
\left( \frac{\beta_0}{0.5} \right)
\left(\frac{M_{\rm SN}}{10 M_{\odot}}\right)^{-3/2}
\left(\frac{E_{\rm SN}}{2\times 10^{51} \erg}\right)^{3/2}
 \\ & \times
\left(\frac{\kappa_{\rm c}}{0.1  \cm^2\g^{-1}}\right)^{-1}
\left(\frac{t_{\rm j,f}}{100 {~\rm d}}\right)^3 \erg \s^{-1}. 
\end{aligned}
\label{eq:Ljetmax}
\end{eqnarray}
\normalsize

I took the rise time to be the end time of the activity phase of the powerful jets, i.e., $t_{\rm rise} \simeq t_{\rm j,f}$, although the rise time might be somewhat longer than $t_{\rm j,f}$ because of photon diffusion time. 
The red line in Fig. \ref{Fig:LSNeSLSNe} is equation (\ref{eq:Ljetmax}) for 
$L_{\rm c}$ as expressed in r-band magnitude $M_{\rm r}= 4.64 - 2.5 \log (L_{\rm c}/L_\odot)$. 

The red line in Fig. \ref{Fig:LSNeSLSNe} is not unique. It rather represents a trend in cases where long-duration jets power the longest LSNe and SLSNe. 
This relation is not unique because of the very large parameter space of jet-ejecta interaction, as I mentioned already in this review. The toy model reflects this large parameter space, but does not include all parameters. For example, the power of the jets is unlikely to be constant as I assumed above. 
The trend that this relation (equation \ref{eq:Ljetmax}) demonstrates is that for massive ejecta of $M_{\rm SN} \ga 10 M_\odot$ the peak luminosity rapidly increases with the duration of the jet activity, $L_c \propto t^{\xi}_{\rm j,f}$ with $\xi \simeq 3$. The massive ejecta might account also for the long-duration fallback accretion that feeds the accretion disk that launches the jets. 
 
Jets might explain also the powering of LSNe and SLSNe that are to the left of the red line. For example, the jet activity phase might be shorter. It is also possible that  the powering of these CCSNe is only by jets launched at the explosion time itself. 
As well, for a given jet-activity duration the peak luminosity can be much higher 
if the ejecta mass is smaller and/or the jets are more energetic ($\epsilon_{E,0}$ is larger). 
In addition, the viewing angle plays a significant role in determining the light curve (e.g., \citealt{AkashiMichaelisSoker2022}). 

\textbf{$\bigstar$ Summary of section \ref{subsec:LongLastingjets}.}
Long lasting jets might account for the general right boundary of most CCSNe-I (hydrogen deficient CCSNe) in the plane of the peak r-band magnitude versus rise time.
Namely, for the LSNe and SLSNe with the longest rise time (beside those to the far right). In some cases strong jets at explosion form bipolar ejecta that with a correct viewing angle might account for some peculiar CCSNe. 
I do not claim that magnetars and/or ejecta-CSM interaction do not add to the powering of these LSNe and SLSNe. My claim is rather that we must include jets when we use magnetar powering, and should consider jets alongside ejecta-CSM interaction. 

\section{Open questions}
\label{sec:Summary}
I list some of the key open questions. 
\begin{itemize}
\item  Can the pre-collapse core convection zones supply sufficient amount of angular momentum fluctuations to lead to the formation of intermittent accretion disk or belt around the newly born NS? Can such stochastic intermittent accretion disks/belts launch sufficiently energetic jittering jets to explode the star? I presented arguments in this review for positive answers to both questions, but there is a need for high-resolution magnetohydrodynamical simulations to confirm my claim. 
\item What is the exact relation and mutual influence of powering by jets and of neutrino heating? 
\item What determines whether the CCSN remnant is a NS or a BH? My view is that pre-collapse rapid core rotation implies inefficient jet feedback mechanism that in turn leads to the formation of a BH. Again, this requires confirmation by high-resolution magnetohydrodynamical simulations. 
\item Are there failed CCSNe? My view as I presented here is that there are no failed supernovae. Even when the remnant is a BH there is a bright explosion driven by energetic jets. Further observational and theoretical studies are needed.  
\item What is the relative role of jets, magnetars, and ejecta-CSM interaction in powering SLSNe? I presented here arguments for that jets supply most of the energy in most cases, but in many cases the two other processes also play roles.  
\item  What are the roles of jets that a companion launches before explosion in shaping the CSM, and how does the binary interaction influence the explosion (like stripping the envelope of the progenitor and/or spinning-up its core)? 
\item What is the role of jets that a companion launches in powering pre-explosion outbursts? 
\item What is the relation of the triggering mechanism of pre-explosion outbursts to the explosion mechanism? I argued that strong convective motion in the pre-collapse core are behind these two mechanisms.
\item  How can we observationally distinguish between peculiar CCSNe and CEJSNe? 
\end{itemize}

\section*{Acknowledgements }
 
I thank Paz Beniamini, Nikolai Chugai, Fabio De Colle, Chris Fryer, Ore Gottlieb, Diego L{\'o}pez-C{\'a}mara, Keiichi Maeda, Salvatore Orlando, Yu Wang,  Yossef Zenati, and an anonymous referee for helpful comments. 
This research was supported by a grant from the Israel Science Foundation (769/20).


\end{document}